\documentclass[preprint,5p,times,twocolumn]{elsarticle}

\usepackage{amssymb}

\usepackage{multirow}

\usepackage[final]{changes}

\usepackage{amsmath}
\usepackage{color}

\usepackage{flushend}

\usepackage[acronym]{glossaries}
\newacronym{dso}{DSO}{Distribution System Operator}
\newacronym{gam}{GAM}{Generalised Additive Model}

\journal{Sustainable Energy, Grids and Networks}

\begin{document}

\begin{frontmatter}



\author[lab1]{Ciaran Gilbert}
\author[lab2]{Jethro Browell\corref{cor1}}
\ead{jethro.browell@glasgow.ac.uk}
\author[lab1]{Bruce Stephen}
\address[lab1]{Department of Electronic and Electrical Engineering, University of Strathclyde, Glasgow, G1 1XQ, UK}
\address[lab2]{School of Mathematics and Statistics, University of Glasgow, G12 8TA, UK}

\cortext[cor1]{Corresponding author}

\title{Probabilistic load forecasting for the low voltage network: forecast fusion and daily peaks}






\begin{abstract}

Short-term forecasts of energy consumption are invaluable for \added{the} operation of energy systems, including low voltage electricity networks. However, network loads are challenging to predict when highly desegregated to small numbers of customers, which may be dominated by individual behaviours rather than the smooth profiles associated with aggregate consumption. Furthermore, distribution networks are challenged almost entirely by peak loads, and tasks such as scheduling storage and/or demand flexibility maybe be driven by predicted peak demand, a feature that is often poorly characterised by general-purpose forecasting methods. Here we propose an approach to predict the timing and level of daily peak demand, and a data fusion procedure for combining conventional and peak forecasts to produce a general-purpose probabilistic forecast with improved performance during peaks. The proposed approach is demonstrated using real smart meter data and a hypothetical low voltage network hierarchy comprising feeders, secondary and primary substations. Fusing state-of-the-art probabilistic load forecasts with peak forecasts is found to improve performance overall, particularly at smart-meter and feeder levels and during peak hours, where improvement in terms of CRPS exceeds 10\%.



\end{abstract}




\end{frontmatter}


\section{Introduction}


Distribution networks are in a state of transition, with their original remit of `fit and forget' now supplanted by the potential of massive increase in utilisation from electrified transport and heat, and two-way power flow from embedded renewable generation. This has led to the emerging necessity of the \gls{dso}, who holds responsibility for balancing power flows under the transmission network\added{ \cite{Mokryani2022TransitionOperator}}. While balancing has been commonplace at transmission-level for decades and the requisite forecasting and dispatch capabilities well understood, there is not a direct translation from transmission to distribution. Going down the voltage levels in power networks makes individual, low diversity, demand behaviours less predictable and hence unsuited to the methods used at transmission and regional levels.
%
Short-term forecasts of load at all levels of the distribution network will be essential to coordinate flexibility services from distributed energy resources.

Load forecasting on the transmission network is a highly active area of research, and has been a mature technology for decades. Research in \replaced{recent}{resent} years has been focused on probabilistic forecasts, which communicate the uncertainty associated with a forecast to end-users. There is a growing appetite for such forecasts in industry, which are now used by both Transmission System Operators (TSOs) and energy traders in operational decision making. Probabilistic load forecasting is extensively reviewed in~\cite{Hong2016ProbabilisticReview}, where the authors highlight the emergence of household level forecasting in the context of hierarchical modelling; an opportunity provided by newly available smart meter datasets. 

\begin{figure}[t]
\centering
\includegraphics[width=\columnwidth]{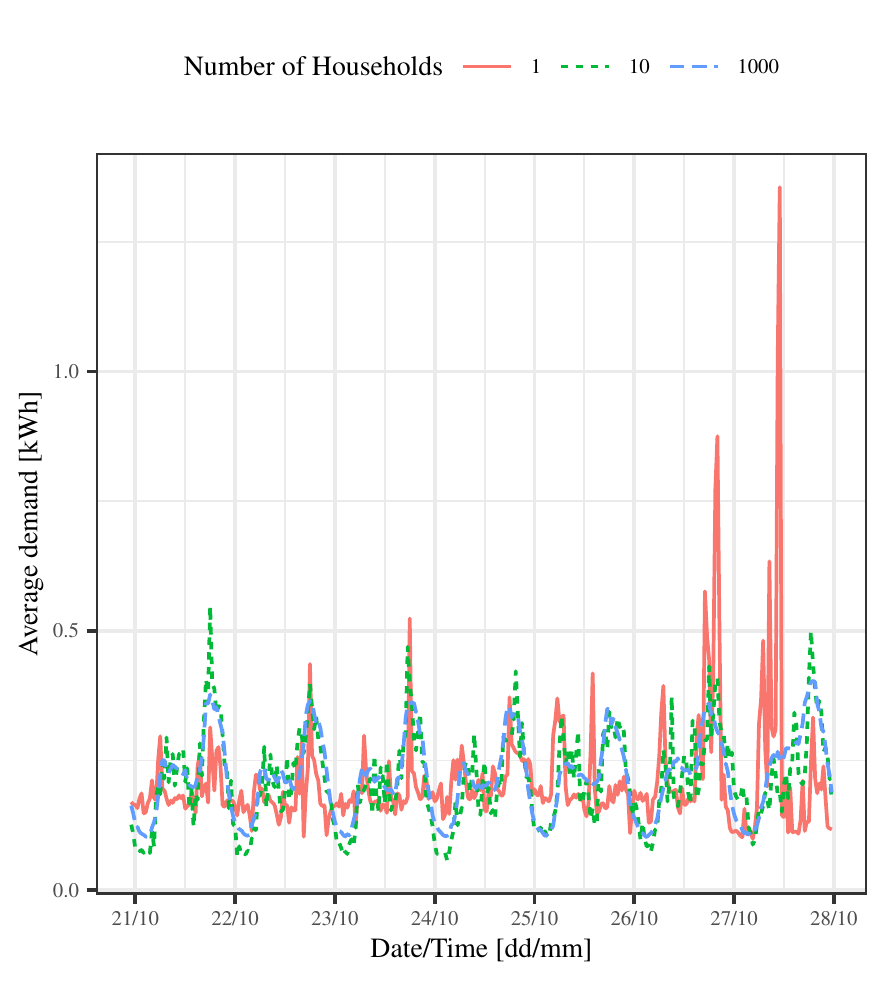}
\caption{Half-hourly demand during one week in 2013 averaged across an increasing number of households from 1 to 1000. Group behaviours become more apparent as the aggregation level increases.}
\label{fig: sm_to_aggregate}
\end{figure}

Forecasting at Low Voltage (LV) levels poses a different challenge to the conventional load forecasting problem at the transmission level. As electricity is aggregated group behaviours emerge which tend to change slowly and are therefore relatively predictable. Disaggregated demand at the household level is much more changeable and influenced by individual behaviours and processes, as shown in Figure~\ref{fig: sm_to_aggregate}. The effect of the signal to noise ratio at the various voltage levels is discussed in much of the following literature where it is suggested that new approaches to forecasting are required and that should be developed with end-use in mind.

Apart from the challenge of the lower signal to noise ratio at the household level, there are also challenges relating to the large number of nodes in LV networks where forecasts may be required, limited coverage of monitoring, data quality, and data privacy. These constraints \replaced{affect}{effect} the applicable methodologies for forecasting; for instance the models must be computationally efficient and the input features may not be location specific. Challenges and opportunities for low voltage forecasts are discussed in an extensive review of the literature\added{ \cite{Zhu2022ReviewSystems,Haben2021ReviewRecommendations}}, where the authors outline recommendations for future research; such as the need for probabilistic forecasting\added{, handling limited observability}, robust forecast verification, and avoiding widespread single-source data-bias in research projects.

The smart meter roll-out in Great Britain and around the globe presents new opportunities in household load forecasting. This area has received the most attention in the literature around LV forecasting, which has mainly focused on deterministic forecasting~\cite{Yildiz2017RecentData,Wang2019ReviewChallenges}. The high penalisation of phase or `timing' errors by traditional point-wise deterministic metrics like mean absolute error is highlighted in~\cite{Haben2014AConsumption}, where a new evaluation measure based on temporal permutation is proposed. The problem highlighted is that typical evaluation metrics tend to reward a smoother forecast on average compared to a forecast that may better represent the underlying process, that misses the precise timing of a sharp increase in demand. Therefore, the importance of predicting peaks in household (and LV) electricity demand has been discussed extensively in the literature \cite{Haben2021ReviewRecommendations}.

The volatility of household demand necessitates a probabilistic approach to forecasting. Univariate probabilistic forecasts are typically communicated as full density forecasts, which are the most flexible for use in decision making, or in the form of multiple discrete quantiles at various probability levels. Interval forecasts with a specific coverage probability are also common~\cite{Bessa2017TowardsIndustry}. In~\cite{Arora2016ForecastingEstimation} density forecasts are obtained using Kernel Density Estimation (KDE), but conditional on information such as time-of-day, with a boundary correction applied to account for the positive nature of demand. Similarly, beta kernels are used in~\cite{Reis2017AGrids} to address the same problem, with a focus on building a scalable forecasting algorithm. Full density forecasts are generated as a benchmark model in~\cite{Wang2019ProbabilisticLSTM}, where the conditional density is assumed to be Gaussian with variance conditional by time-of-day; this is compared to non-parametric forecasts produced using an LSTM network for quantile regression, which outperforms the conditional density approach for the quantiles considered. A quantile regression approach based on boosting with additive models is demonstrated in~\cite{BenTaieb2016ForecastingRegression}, where the additive models are flexible and benefit from automatic feature selection by nature of the component-wise boosting procedure. Importantly, the quantile forecasts at the smart meter level are shown to be more skilful than an advanced parametric approach based on the Gaussian distribution.
\added{Bernstein polynomials have also been proposed for producing non-parametric density forecasts in \cite{Arpogaus2022Short-TermFlows} which show improvement over Gaussian and Gaussian mixture density forecasts.}
Finally, multivariate forecasts are generated for a hierarchy of smart meters in~\cite{Taieb2020HierarchicalData}, where a coherency constraint is placed on the samples of the multivariate distribution, i.e. lower levels must sum to higher levels of the hierarchy.
\added{This literature may give the impression that only non-parametric densities are suitable for household load forecasting, however, there have been no studies examining alternatives to the Gaussian distribution until the present work.}


Some of the works already discussed (e.g.~\cite{BenTaieb2016ForecastingRegression, Taieb2020HierarchicalData}) also forecast low- and medium-voltage levels, i.e. feeder and substation load. however, the networks are hypothetical, in that they are generated from aggregated smart meter data. As discussed in~\cite{Haben2021ReviewRecommendations}, this necessarily excludes important elements on the distribution network such as commercial loads, street lighting, embedded generation, electrical losses, dependencies between customers, etc. However, there is a \replaced{shortage}{lack} of open-source \deleted{available}
data sets \replaced{that}{which} contain a satisfactory amount of nodes due to a lack of widespread monitoring. An alternative option is to generate a synthetic LV dataset from smart meter data and basic information on the local network architecture~\cite{Telford2021DirichletObservability}. In~\cite{Haben2019ShortLevel} several probabilistic methods (quantile regression, KDE) are evaluated using a dataset comprising of 100 real LV feeders where there was no clear best forecaster at all feeders, however, autoregressive type models performed well, and (forecast) temperature was shown to have negligible influence on the forecast skill.

Peak demand is typically the limiting factor in the capacity of distribution networks, set by the maximum power a cable of transformer can handle. Additionally, for a lot of flexibility applications the main goal is to reduce, flatten, or shift the daily peak demand in the LV network. Therefore, day-ahead forecasts of the daily maximum at the different nodes are valuable from both utilities perspective (e.g. in setting dynamic prices) and the consumers perspective (e.g. for scheduling battery or EV charging) \cite{Mokryani2022TransitionOperator}. The level of the daily peak is only half of the issue however, forecasting the time-of-peak is also relevant. There is little published in this specific area for the LV network to the best of our knowledge; related work~\cite{Jacob2020ForecastingPeaks} focuses on Extreme Value Theory and peaks over a defined threshold, which are by definition rare.

In this paper, we consider a four-level hierarchy: a primary substation (33kV--11kV), secondary substations (11kV--4151kV), feeders (415kV), and households (230V, single phase). Methods for generating sharp and calibrated probabilistic forecasts of demand for the day-ahead are \replaced{developed, including non-Gaussian parametric density forecasting at the household level, which is innovative in and of itself}{described}. We also investigate probabilistic daily peak forecasting, as in the daily maximum average energy demand, in terms of both peak intensity and timing. Finally, a method for combining (or blending) the daily bivariate (level and timing) peak forecasts and the half-hourly demand forecasts is described, which is termed forecast fusion. \added{While methods for combining forecasts of varying spatial and/or temporal resolution have been proposed, combining forecasts of related yet distinct quantities has not been explored before, to the best of our knowledge.} The case study presented is based on a hypothetical network generated using the Low Carbon London dataset, and the proposed forecasts are robustly verified against benchmark models.

Probabilistic load forecasting is introduced in Section \ref{sec:prob_forecasting}, followed by the concept of forecast fusion in Section \ref{sec:forecast_fusion}. State-of-the-art methods for conventional day-ahead load forecasting are presented in Section \ref{sec:Halfhourly}, followed by proposed methods for daily peak intensity and timing forecasting in \ref{sec:peak_intensity} and \ref{sec:peak_timing}, respectively. An extensive, fully reproducible, case study based on the Low Carbon London dataset \cite{AMIDiNe_LV_SupMat2022} and a hypothetical LV network is then presented where forecasts for load at household, feeder, secondary substation and primary substation levels are analysed in detail. Finally, brief conclusions are drawn in Section \ref{sec:conculsions}.

\section{Probabilistic Forecasting}
\label{sec:prob_forecasting}

In this section the probabilistic forecasting framework is formalised and the flexible statistical learning framework employed to generate forecasts throughout the work is introduced. We are typically interested in the the predictive density or cumulative distribution function (CDF) of random variable $Y_t$ at time $t$. The predictive CDF is defined as
\begin{equation}
	\hat{F}_t(y_t)= P(Y_t\leq y_t) 
\end{equation}
where $\hat{F}$ is a strictly increasing function. Here we consider time in half-hour periods, as this is the resolution of electricity metering in Great Britain, but other this is not a restriction on the methodology. A density forecast provides maximum flexibility as quantiles or intervals can easily be extracted from the forecast, and no need to approximation are necessary, such as interpolating between quantiles. Additionally, the full distribution may be described by a smaller number of parameters, which may be functions of explanatory variables. One drawback is that a suitable conditional parametric family must be found for the forecast. Kernel density estimation provides a non-parametric alternative but can be less flexible and more computationally demanding and therefore less scalable --- critical for \glspl{dso} where the number of assets is large.

\subsection{Generalised Additive Models for Location, Scale, and Shape}

Generalised Additive Models for Location, Scale, and Shape (GAMLSS)~\cite{Rigby2005GeneralizedShape}, are semi-parametric models. This is because a parametric distribution is assumed for the target variable, and the parameters that define the assumed distribution may depend on non-parametric smooth functions of explanatory variables. The framework is an extension of the more familiar \gls{gam}~\cite{Hastie2009ThePrediction}, such that any parameter of the distribution can be a function of input features, not just the conditional mean.

If we have observations $y$, in this case demand at a particular location on the LV network, the conditional density typically $f(y|\boldsymbol{\theta})$ depends on up to four parameters; these are the location ($\theta_1$), scale ($\theta_2$), and shape parameters ($\theta_3, \theta_4$). An additive regression model 
is generated for each distribution parameter $\theta_i$ for $i=1,\ldots,4$. Let $\boldsymbol{x}_i$ be the pool of $N_i$ input features in the sub-model for $\theta_i$, and $g_i(\cdot)$ the link function, then the model formulation of a GAMLSS is
\begin{equation}
    g_i(\theta_i) = \beta_{0,\theta_i} + \sum_{n=1}^{N_i}f_{n,\theta_i}(x_{i,n}), \quad i = 1,\ldots,4
\end{equation}
where the function $f_{n,\theta_i}$ is the effect of explanatory variable $n$ on the distribution parameter $\theta_i$, which can be linear or non-linear functions, such as penalised smoothing splines, linear coefficients, surfaces, etc; $\beta_{0,\theta_i}$ are the intercepts of each sub-model. These models may be estimated numerically using a combination of maximum likelihood, and successive back-fitting of the predictor functions for each parameter~\cite{Rigby2005GeneralizedShape}.

\begin{figure}[t]
\centering
\includegraphics[width=\columnwidth]{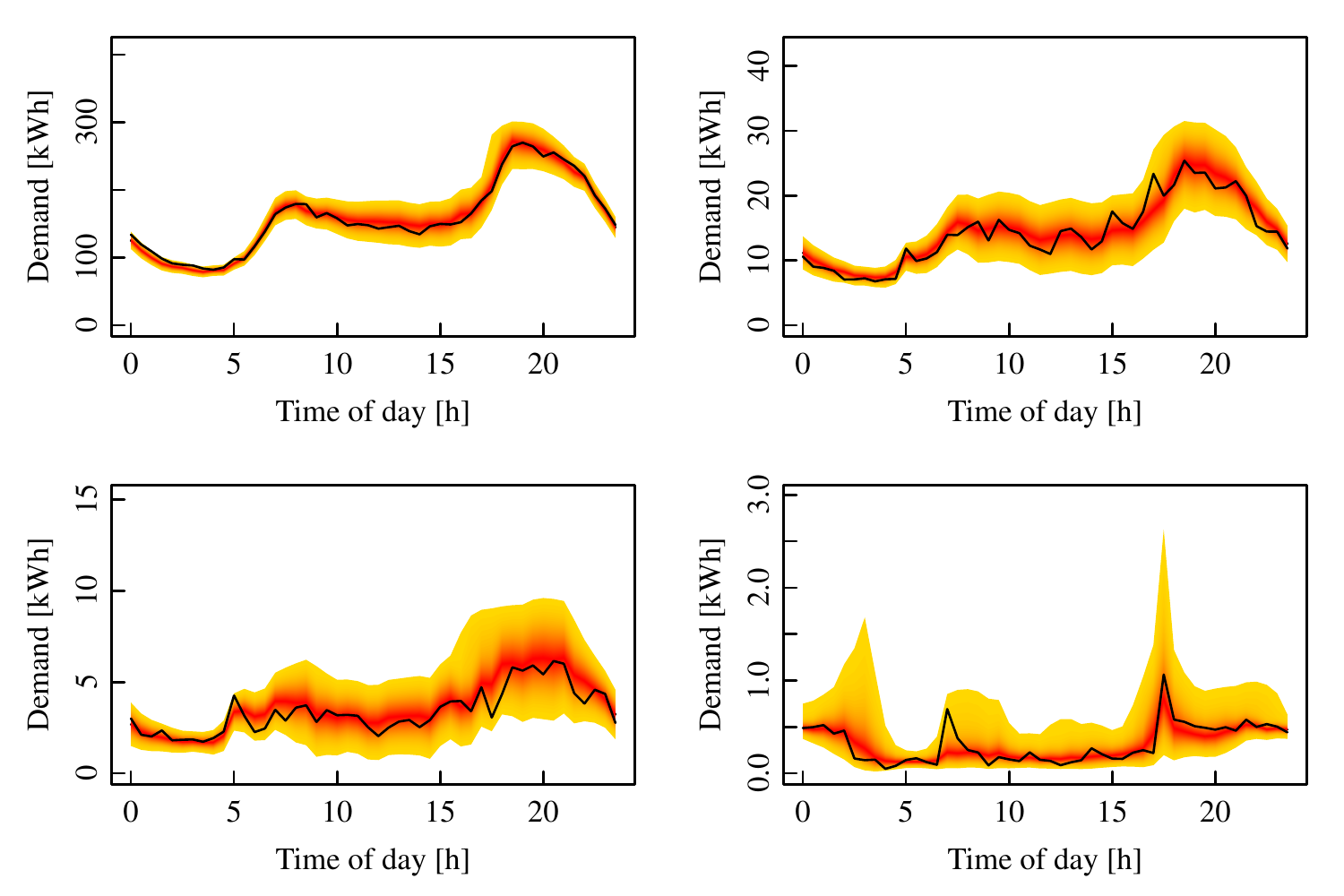}
\caption{Day-ahead example probabilistic forecasts as fan plots for the primary substation (\texttt{ps1} top left), a secondary substation (\texttt{ss1} top right), a feeder (\texttt{ss1\textunderscore fdr1}, bottom left), and a household (\texttt{N1174}) on 17-10-2013. The widest and lightest coloured interval has a coverage probability of 98\% and the measurement is overlaid in black for reference.}
\label{fig: exfc}
\end{figure}

In Figure~\ref{fig: exfc} four example density forecasting models are visualised in a fan plot, where probability intervals are extracted from the conditional distribution at each lead time. Load forecasts \deleted{forecasts} are shown at different levels of aggregation. In general, forecast uncertainty is greater the further load is disaggregated. In particular, the possibility of large peaks is clearly quantified by high quantiles of the household-level forecast, which would not be captured by point forecasts.

Importantly, demand can approach zero in households although in our case study framework cannot be negative (net-demand, demand less embedded generation, is reserved for future work). This removes some parametric families for the forecasts from consideration, such as the Gaussian distribution. In Figure~\ref{fig: exfc}, the Generalised Beta Prime distribution~\cite{Stasinopoulos2007GeneralizedR} is used for disaggregate demand and the Gaussian distribution is used for the three other aggregated levels. This small example is indicative of the approach throughout, which was to model the aggregate and household levels in the network distinctly; the forecasting models at the household level have to be simple, computationally efficient, and be suitable for right-skewed data.

\section{Forecast Fusion}
\label{sec:forecast_fusion}

In this paper, we hypothesise that better forecast skill can be achieved by fusing a bespoke forecast of the daily peaks in demand, in terms of both the timing and intensity of the event, with a state-of-the-art half-hourly resolution forecast, both produced one day-ahead. This technique is very similar to forecast combination (or blending, expert mixtures, etc.), but distinct in that we are combining forecasts of two different types: the demand for a particular half-hour and day $y_{d,h}$ and the daily peak in demand $y^{(p)}_{d}$. A forecast of the timing of the peak, i.e. the number of half-hours from midnight until the daily peak $h^{(p)}_{d}$, is also used in the method.

Consider the fusion of the two forecasts as a linear combination of the two distribution functions, 
\begin{equation}
	\Tilde{F}_{d,h}(y)= (1-w_{d,h}) \hat{F}_{d,h}(y) + w_{d,h}\hat{F}^{(p)}_{d}(y)
\end{equation}
with weights $w_{d,h}$ that may also be forecasts,or otherwise chosen, for each half-hour. This formulation respects the fact that a probabilistic forecast of the daily peak intensity (maximum value) is not sufficient in most applications and that the timing of the peak \added{is} also relevant. In this paper the weights are derived from probabilistic forecasts of the time-of-peak. If we define the random variable $H=\{1,2,...,48\}$ to be the number of half-hours to each daily maximum in demand from midnight $h^{(p)}_{d}$, then the weights are found via a forecast probability mass function
\begin{equation}
    w_{d,h} = \hat{f}_{d,h}(h) = P(H = h)
\end{equation}
since forecasts are typically issued for discrete blocks of energy. Therefore, estimating the weights is re-framed into a discrete time-to-event prediction problem. Note that the sum of the probabilities $\hat{f}_{d,h}(h)$ over the total number of blocks in each day should \replaced{be}{sum to} 1. If the conventional forecast were a prediction of load conditional on there not being a peak at $d,h$, this could be interpreted as a statement of the law of total probability. While it may be possible to produce such a forecast, we choose to proceed using conventional day-ahead forecasts so that the fusion method is as applicable as possible,

Forecast fusion is related to techniques found in probabilistic forecast combination~\cite{Browell2020QuantileCompetition, Baran2018CombiningForecasts}, blending~\cite{Kober2012BlendingForecasts}, expert mixtures~\cite{Capezza2021AdditiveForecasting}, linear pools~\cite{Gneiting2013CombiningDistributions, Ranjan2010CombiningForecasts}, and so on. Empirically it was found that important concepts in forecast combination, such as re-calibration of a linearly combined forecasts, were not necessary in the following case study. A beta-transformation of the combined forecast, following~\cite{Gneiting2013CombiningDistributions}, reduced the forecast skill in cross-validation; further work on forecast combination may yield benefits here. However, all the aforementioned methods are based on combining forecasts of the same type (e.g. hourly resolution, day-ahead forecasts), hence the distinct terminology here, which highlights the similarity with the broader topic of data fusion.

Data fusion can be defined as the combination of multiple sources to obtain improved information in terms of quality, expense, and/or relevancy~\cite{Castanedo2013ATechniques}. This is typically applied to combining data from multiple sensors and sources. In~\cite{Abrahart2002Multi-modelCatchments} multi-model forecast combination is discussed in terms of data fusion, however, the method entails combining forecasts of the same variable much like the literature discussed previously.
In this work, we propose the fusion of conventional load forecasts with forecasts of peak load. 
These two forecasts exist in different temporal domains, hence this is a fusion problem not the usual practice of forecast combination or blending. We hypothesise that this will improve forecast skill overall as well as at peaks specifically, similar to how reconciliation of temporal hierarchies have been found to improve the skill across the different temporal domains~\cite{Bergsteinsson2021HeatHierarchies}.

\section{Day-ahead Load Forecasting}
\label{sec:day-ahead_forecasting}

In this section\added{, the} approach for generating the day-ahead probabilistic forecasts of half-hourly load, peak intensity and peak timing are detailed. Throughout, different approaches are used depending on the level of aggregation in the LV network, including model specifications and input features. The household level is treated as one group, and feeder level and above as the other, referred to as `aggregated' levels. This prevents the framework from becoming too complex but is not a strict constraint. Common to all base probabilistic forecasts generated (the half-hourly forecasts $\hat{F}_{d,h}(y)$, the daily peak intensity forecasts $\hat{F}^{(p)}_{d}(y)$, and the peak timing forecasts $\hat{f}_{d,h}(h)$) is the \gls{gam} framework, where both half-hourly and peak demand forecasts utilise GAMLSS.

\subsection{Half-hourly Forecasts}
\label{sec:Halfhourly}

The half-hourly forecasts at the {aggregated} levels of the LV network are described by $f_{d,h}(y|\theta_{1_{d,h}}; \theta_{2_{d,h}})$ where we assume the conditional distribution is Gaussian. The model is formulated as follows at each aggregate node
\begin{equation}
\begin{aligned}
g_1(\theta_{1_{d,h}}) = &~\beta_{0} + \beta_{1}y_{d-1,h} + \beta_{2}y_{d-7,h} + \beta_{3}y^{(p)}_{d-1} + \\ &
\sum_{j=1}^{48} \alpha_{j}H_{j}(h)y_{d-1,h} + \sum_{j=1}^{48} \gamma_{j}H_{j}(h)y^{(p)}_{d-1} + \\ &
f_{pvc}(h,D(d)) + f_{pbc}(d)\\
\end{aligned}
\end{equation}
and for the scale parameter, the formula is reduced to only depend on time-of-day for robustness
\begin{equation}
g_2(\theta_{2_{d,h}}) = \beta_{0} + f_{pb}(h)
\end{equation}
where $f_{pvc}(\cdot)$ is a varying coefficient penalised spline, $f_{pbc}(\cdot)$ is a cyclic penalised spline, and $f_{pb}(\cdot)$ is a penalised spline. There are two dummy variables for each half of the day $H_j(h)$, and period of the week $D(d)$ which is split into day type (Weekday, Saturday, and Sunday). So the load forecast is dependent on lags of the demand, a lag of peak demand, and interactions between yesterday's lags for each half-hour of the day. We also model the diurnal trend via the varying coefficient spline which changes according to day-type. Finally the annual seasonality is included, although in practice this spline is constrained to be very smooth to prevent interpolation of the data (in the case study we only have one year of data). Other formulations were tested; a full exploratory analysis can be found in the supplementary material~\cite{AMIDiNe_LV_SupMat2022}. However, this model formulation produced skilful forecasts relative to the benchmarks averaged over all time periods as well as during daily peaks, thanks to the interaction terms and the simple formula for the scale parameter.

Due to the complexity and sheer number of households, the half-hourly forecast models for the {household} level have to be more simple and robust than those at the aggregate levels. This is to prevent over-fitting and for computational efficiency. They are given by $f_{d,h}(y|\theta_{1_{d,h}}; \theta_{2_{d,h}}; \theta_{3_{d,h}}; \theta_{4_{d,h}})$ where we assume the conditional distribution follows the Generalised Beta Prime distribution. The model is formulated as
\begin{equation}
\begin{aligned}
g_1(\theta_{1_{d,h}}) = &  \beta_{0} + \beta_{1}y_{d-1,h} + \beta_{2}y_{d-7,h} + \sum_{j=1}^{7} \alpha_{j}D_{j}(d) + \\& 
f_{pb}(h) + f_{pbc}(d)\\
\end{aligned}
\end{equation}
and for the scale parameter, the formula is similar to the aggregated levels
\begin{equation}
g_2(\theta_{2_{d,h}}) = \beta_{0} + f_{pb}(h)
\end{equation}
and the two shape parameters of the distribution are constants to be estimated. Here the dummy variable $D_j(d)$ is for each of the 7 periods of the week.

\subsection{Daily Peak Intensity}
\label{sec:peak_intensity}

\begin{figure}[t]
\centering
\includegraphics[width=\columnwidth]{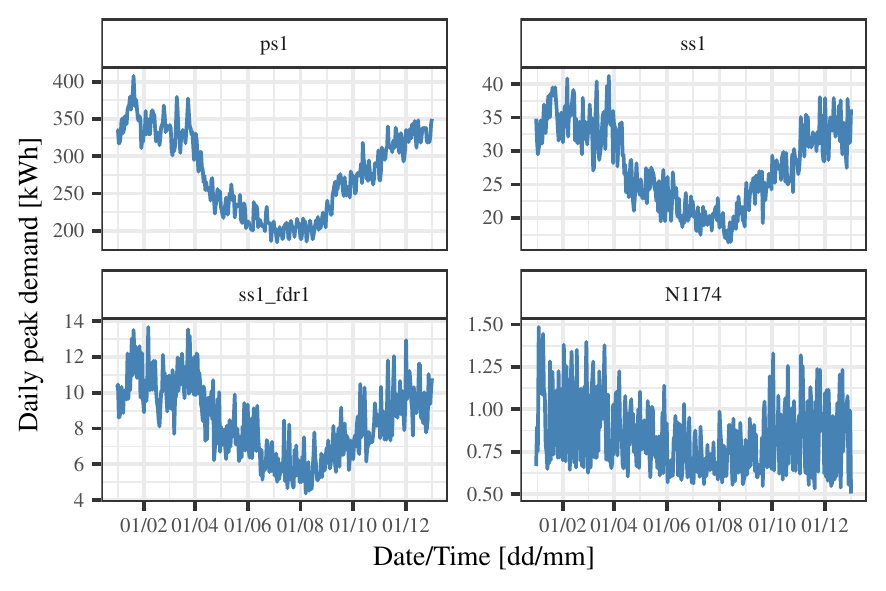}
\caption{Example time series of the daily peak intensity for
primary substation (\texttt{ps1}), a secondary substation (\texttt{ss1}), a feeder (\texttt{ss1\textunderscore fdr1}), and a household (\texttt{N1174}), from the hypothetical LV network. The peak demand shows seasonality at the aggregated levels, but again is more volatile at this household.}
\label{fig: peakts}
\end{figure}

\begin{figure}[t]
\centering
\includegraphics[width=\columnwidth]{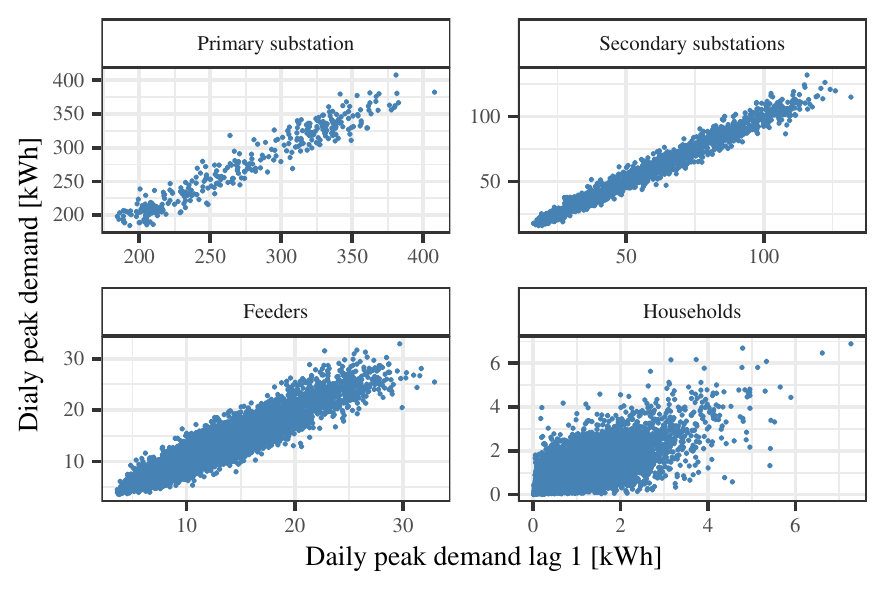}
\caption{Lag dependency plots of the the daily peak intensity at four levels in a hypothetical LV network. This includes all nodes at each level and shows the motivation for using autoregressive based models for peak intensity forecasting, especially at the aggregate levels.}
\label{fig: lagdep}
\end{figure}

For the daily peak intensity forecasting, data exploration revealed seasonal trends and a high correlation in the lag dependency variables, as shown in Figures~\ref{fig: peakts} and~\ref{fig: lagdep} respectively, in the hypothetical LV network. Albeit the strength of the relationships are much weaker at the household level, again due to the low signal to noise ratio. An important consideration when creating the forecasting models here is the reduced size of the data set, since there is only one data point per day. Therefore, we reduce the number of features and categories in each formulation compared to half-hourly forecasting.

The peak intensity forecasts at the {aggregated} levels of the LV network are described by $f^{(p)}_{d}(y|\theta_{1_{d}}; \theta_{2_{d}})$ where we assume the conditional distribution is Gaussian. The model is formulated as follows at each aggregate node
\begin{equation}
\begin{aligned}
g_1(\theta_{1_{d}}) = &  \beta_{0} + \beta_{1}y^{(p)}_{d-1} + \beta_{2}y^{(p)}_{d-7} + \beta_{3}\sigma_{y_{d-1}} + f_{pbc}(d) + \\& 
\beta_{4}D_{1}(d) + \beta_{5}D_{2}(d)\\
\end{aligned}
\end{equation}
and for the scale parameter, the formula is
\begin{equation}
g_2(\theta_{2_{d}}) = \beta_{0} + \beta_{1}D_{1}(d) + \beta_{2}D_{2}(d) + \beta_{3}\sigma_{y_{d-1}}
\end{equation}
where period of the week $D_i(d)$ is now reduced simply to either weekday or weekend categories, and $\sigma_{y_{d-1}}$ is yesterday's standard deviation of the half-hourly demand time-series.

An important feature at the {household} level for model robustness is the `empty house' feature $I(d)$. We include lags of this variable, which defined an empty house by look at run-lengths of the daily standard deviation of the half-hourly demand. If this value dropped to approximately zero for a period of at least 7 days then the empty house feature is active. Note that the household had to have at least 30 days of emptiness for the feature to be included. This is because we want to identify houses here that are regularly empty for a reasonable period of time, rather than capture things such as holidays, public holidays, etc. which are reserved for future work. So, the daily peak intensity forecasts at the household level are described by $f_{d}(y|\theta_{1_{d}}; \theta_{2_{d}}; \theta_{3_{d}}; \theta_{4_{d}})$ where we assume the conditional distribution follows the Generalised Beta Prime distribution. The model is formulated as follows at each household
\begin{equation}
\begin{aligned}
g_1(\theta_{1_{d}}) = &  \beta_{0} + \beta_{1}y^{(p)}_{d-1} + \beta_{2}y^{(p)}_{d-7} + f_{pb}(d) + \\& 
\beta_{3}D_{1}(d) + \beta_{4}D_{2}(d) + \beta_{5}I(d-1)\\
\end{aligned}
\end{equation}
and for $g_2(\theta_{2_{d}}) = \beta_{0} + \beta_{1}I(d-1)$, $g_3(\theta_{3_{d}}) = \beta_{0} + \beta_{1}I(d-1)$, the formula for the fourth moment of the distribution is kept constant. So the scale and shape parameters are only dependent on the empty house feature to make the forecasts more robust to overfitting.


\subsection{Daily Peak Timing}
\label{sec:peak_timing}

\begin{figure}[t]
\centering
\includegraphics[width=\columnwidth]{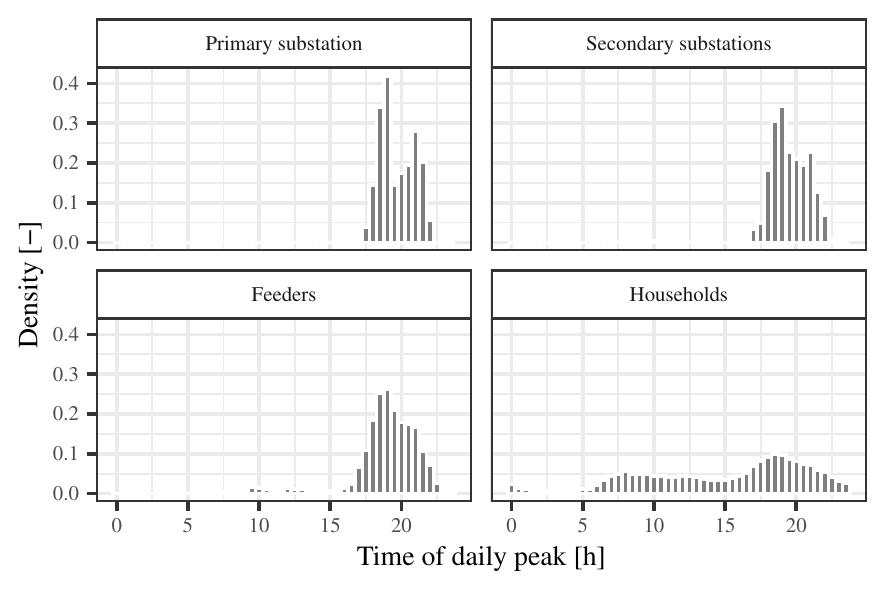}
\caption{Histograms of the daily peak timing at the four different aggregations of the hypothetical LV network. The peak timing becomes more dispersed at the lower aggregations on the network. At the primary substation level the peak timing during the analysis was consistently in the evening, except one data point which corresponds to the `turkey peak' during Christmas day.}
\label{fig: peakt_hist}
\end{figure}

A key component of the forecast fusion method is the weighting. In this study we choose to forecast the weights by defining them as the probability of the peak demand timing over the discrete blocks of energy in a day. Histograms showing the distribution of the peak timing at different levels on a hypothetical LV network are shown in Figure~\ref{fig: peakt_hist}, which shows that as expected the time of the daily peaks become more variable as demand becomes disaggregate. The timing of the peak demand, specially at the higher aggregations, is dependent on the time-of-year; it is widely understood during the winter the peak daily demand tends to be earlier in the evening (and the level of the peak becomes higher, an example of which is shown in Figure~\ref{fig: peakts}) on the GB network. However, complex seasonal interactions were observed in the time-of-peak data, especially at the feeder and secondary substation levels of aggregation.

The process is framed as a discrete time-to-event problem, where with suitable transformations of the time-of-daily peak time series $h^{(p)}_{d}$, a \gls{gam} framework can be applied to generate forecasts~\cite{Bender2018AAnalysis}. This means we can leverage the powerful smoothing capabilities of a \gls{gam} to capture complex seasonal interactions between input features at the daily peak timing. The framework is relatively unique in terms of time-to-event or survival analysis, in that due to the framework there must be an event (i.e. peak) for every subject (i.e. day) for each experiment (i.e. node), and the domain of the peak timing in the analysis is $h^{(p)}_{d} \in \{1, 2, ... , 48\}$.

\begin{figure}[t]
\centering
\includegraphics[width=\columnwidth]{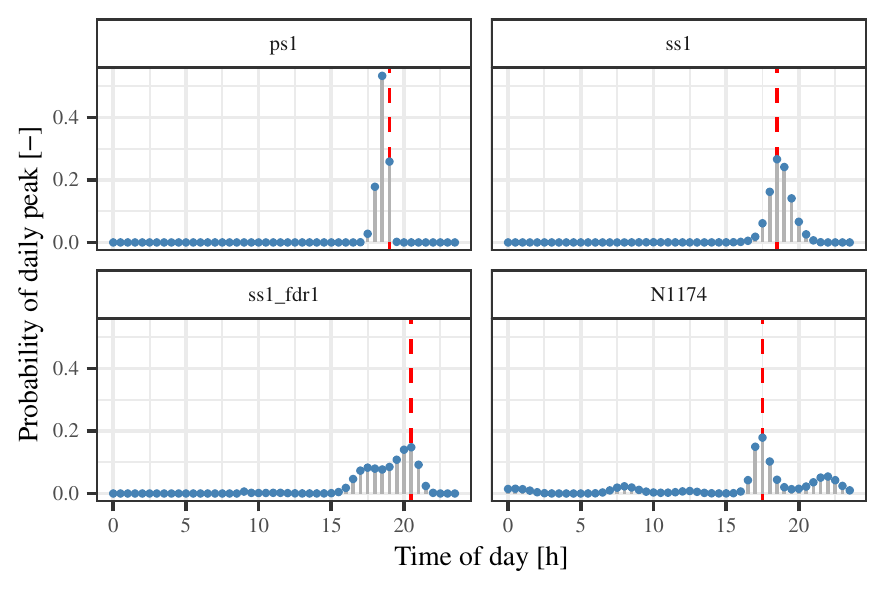}
\caption{Day-ahead example probability forecasts for the time of the daily peak at the same four nodes as Figure~\ref{fig: exfc} and for the same day. The measurement is represented by the dashed red vertical line in each panel for reference. These forecasts correspond to the weights used in the forecast fusion for each discrete time period of this day.}
\label{fig: exfc_peakt}
\end{figure}

The discrete hazard function is the conditional probability at time interval $h$
\begin{equation}
\lambda(h|x) = P(H = h | H \geq h, x) 
\end{equation}
which gives the conditional probability of a peak in interval $h$ given that the peak happens at time $H\ge h$. To model this hazard rate, we use a time varying linear predictor with a \gls{gam} framework
\begin{equation}
\lambda(h|x) = g^{-1}\bigg[\beta_{0h} + \sum_{n=1}^{N}f_{n}(x_{nh})\bigg] 
\end{equation}
where the link function here $g(\cdot)$ is the logit link. In this case for the {aggregated} levels in this case we use a tensor interaction term for the period of day and day of year as the input features as well as a dummy variable for the day type (weekday/weekend). At the {household} level a more simple approach is needed for computational efficiency; two separate smooth splines for each of the input features are used. The discrete survival function is defined as
\begin{equation}
S(h|x) = P(H > h|x) = \prod_{s = 1}^{h}(1-\lambda(h|x))
\end{equation}
from which the discrete cumulative distribution function is found $F(h) = 1-S(h)$, and then the probability mass function for each day $f_{h}(h) = P(H = h)$ is easily calculated. For estimating the regression model, an indicator variable for each daily recorded value of $h^{(p)}$ is defined as the target variable which is zero for each discrete block of time until the event is recorded where the indicator variable is 1. For more information on the data transformation the reader is referred to~\cite{Fahrmeir1997DiscreteModels,Tutz2004FlexibleEffects}. Note that the final peak timing probability forecast could similarly be obtained via a simple logistic regression/classification method, but the discrete time-to-event setting gives the proposed method a stronger theoretical foundation. In Figure~\ref{fig: exfc_peakt} four probability forecasts are shown for the same LV nodes as Figure~\ref{fig: exfc} and for the same day; this constitutes the weights $w_{d,h}$, i.e. the last component required to fuse the half-hourly and daily peak intensity forecasts together.

\section{Case Study}
\label{sec:case_study}

The methodology is tested using data from the Low Carbon London trial~\cite{Tindemans2016Low2013}. Households which are on a variable price tariff are first removed, and then households which have regular communication issues and/or suspect data are removed, and finally only households which have a complete record of measurements during 2013 are retained. This represents quite a strict data cleaning process in which we are left with 742 smart meters; however, addressing challenges due to missing-data and forecasting dynamic price tariff households are beyond the scope of this study. Data from this experiment is anonymised which means location specific effects, such as temperature, are necessarily excluded from the analysis.

\begin{figure}[t]
\centering
\includegraphics[width=\columnwidth]{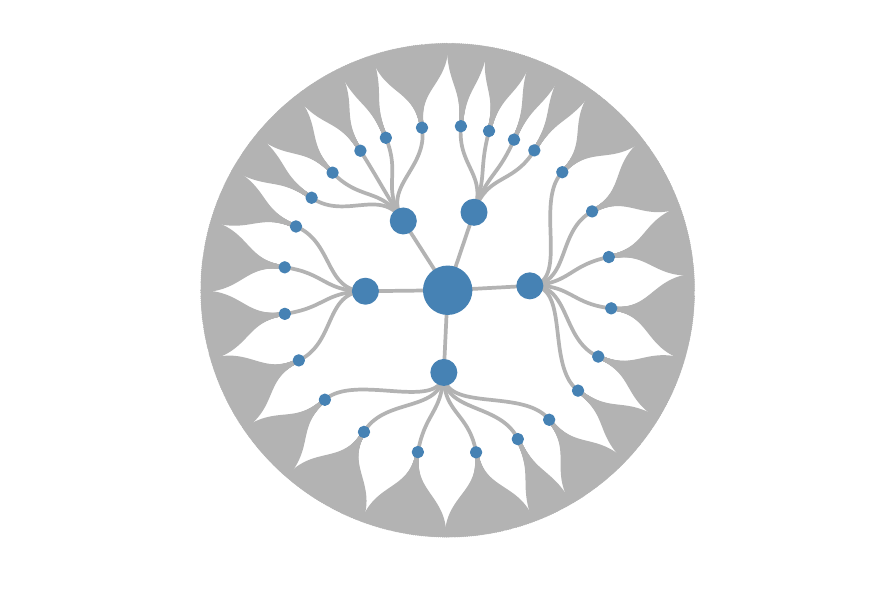}
\caption{Diagram of the hypothetical LV network hierarchy used in the case study where the central node represents the primary substation and so on down to the household level at the end nodes. The size of the primary substation is small compared to reality in the GB distribution network due to the limited number of smart meters used. However, as shown in Figure~\ref{fig: sm_to_aggregate}, group behaviours begin to emerge quickly at high levels of aggregation.}
\label{fig: lvnet}
\end{figure}

The households are sampled (without replacement) to create a hypothetical LV network, albeit given the amount of the available smart meters, the primary substation level (i.e. the top aggregation) is smaller than perhaps you would find in practice. However, as shown in Figure~\ref{fig: sm_to_aggregate}, group behaviours begin to emerge quickly. The sampling process was configured whereby each secondary substation comprises of between 4-7 feeders and each feeder contains around 16-45 smart meters. The hypothetical LV network is illustrated in Figure~\ref{fig: lvnet} showing all the different levels of aggregation. As discussed in~\cite{Haben2021ReviewRecommendations} there is a severe lack of real open-access LV network data available. Therefore, this approach is effectively a compromise because we implicitly cannot account for street furniture, embedded generation, and potential correlations between nodes in a real life network.

For the regression problems, the coverage of the dataset is January 2013 to December 2013 inclusive. To compare the forecast performance, each month is partitioned into thee blocks of approximately 10 days, depending on month. The first two blocks of every month compromise the training data, on which cross-validation is also carried out to generate out of sample forecasts covering the full year. The last block in each month constitutes the testing data, i.e. not used anywhere for model estimation or tuning. All methods are implemented in R~\cite{RCoreTeam20202020R:Computing.} using the package ProbCast~\cite{Browell2020ProbCast:Forecasts}, developed by the authors for the modelling, evaluation, and visualisation of probabilistic forecasts. The wrapper functions for GAMLSS for the regression models~\cite{Rigby2005GeneralizedShape} are used extensively here.

A two pronged approach is necessary to assess \deleted{For} probabilistic forecasts performance; calibration (or reliability) is \replaced{the necessary condition that predicted probabilities are unbiased}{essentially a measure of probabilistic bias}, and sharpness is a measure of \deleted{the accuracy of a probabilistic} forecast uncertainty, i.e. the spread of the \added{predictive} distribution. \replaced{Combined, these}{that} allow for the ranking of competing forecasting \added{methods by sharpness} subject to calibration~\cite{Gneiting2007ProbabilisticSharpness}. For the full predictive cumulative distributions in this case study, the Continuous Ranked Probability Score (CRPS)~\cite{Gneiting2007ProbabilisticSharpness,Jordan2019EvaluatingScoringRules} is used to measure both sharpness and calibration
\begin{equation}
\label{eq: crps}
	 \text{crps} = \frac{1}{N} \sum_{t=1}^{N} \int_{-\infty}^\infty \{\hat{F}_t(y)- \mathbf{1}(y \geq y_t)\}^2dy
\end{equation}
where $\mathbf{1}(\cdot)$ is the indicator function. The CRPS for a single forecast observation pair is therefore the area between the squared difference of the forecast and observation CDF, where the latter is a step-function from 0 to 1 at the observed value. For the discrete probability forecasts of the peak timing, the Ranked Probability Score (RPS) is used for verification, which is the discrete form of CRPS~\cite{Wilks2019ForecastVerification}.

The Probability Integral Transform (PIT) histogram is used to verify the calibration of the forecasts
\begin{equation}
	  u_t= \hat{F}(y_t)
\end{equation}
whereby, if the forecasts are well calibrated and the sample is sufficiently large, $u \sim \mathcal{U}(0,1)$, which is  inspected visually via a histogram with a certain number of (typically 20~\cite{Wilks2019ForecastVerification}) bins. One limitation with this approach is that is becomes time consuming to verify all nodes in a large network. So, in this case the calibration checked within grouped network levels (e.g. smart meters, feeders, etc.). Reliability diagrams~\cite{Messner2020EvaluationView} are used as an alternative to check the calibration of quantiles of the distribution at individual nodes in the supplementary material~\cite{AMIDiNe_LV_SupMat2022}. The calibration of the discrete probability forecasts is not presented here for brevity, but are shown in the supplementary material via multi-category reliability diagrams~\cite{AMIDiNe_LV_SupMat2022,Hamill1997ReliabilityForecasts}.

\subsection{Benchmarks \& Skill Scores}

A set of benchmark models have been chosen for case study, with models chosen depending on the forecasting task and level in the LV network. There a limited literature on benchmarks for probabilistic forecasts in general, and none established for LV load forecasting.

For the half-hourly forecasts a  simple auto-regressive model is used and implemented in the GAMLSS framework for the {aggregated} levels
\begin{equation}
g_1(\theta_{1_{d,h}}) = \beta_{0} + \beta_{1}y_{d-1,h} + \beta_{2}y_{d-7,h} + f_{pb}(h)
\end{equation}
with and the scale parameter and distribution family the matching the proposed advanced model. At the {household} level three benchmark models are employed; two are based on Kernel Density Estimation (KDE) using zero-truncated  Guassian kernels. In the first separate KDE models are defined for each for each half-hour of the day, and in the second separate KDE models are defined for each half-hour and day type (weekday, Saturday, Sunday). Both of these are similar to a previously published method~\cite{Arora2016ForecastingEstimation}. The third is a simplified version of the GAMLSS model used at the household level, with the same Generalised Beta Prime family, where only the location and scale parameters vary smoothly with time-of-day.

For the peak intensity forecasts a very simple autoregressive based GAMLSS model is used for the {aggregated} levels, with the Gaussian conditional distribution and formula
\begin{equation}
\label{eq: intb}
g_1(\theta_{1_{d}}) = \beta_{0} + \beta_{1}y^{(p)}_{d-1} + \beta_{2}y^{(p)}_{d-7} 
\end{equation}
where the scale parameter is a constant. At the {household} level, again three benchmark models are employed. The first is a very simple unconditional KDE estimate, the second is a KDE estimate conditional on day type (weekday/weekend), and finally a simple location-only autoregressive GAMLSS model, i.e. the same as Equation~\eqref{eq: intb} except using the Generalised Beta Prime family as the conditional distribution.

For the peak timing probability forecasts the same simple seasonal climatology model is used at all points of the network. This is a competitive benchmark because seasonality is the only effect used in our more advanced model.

Forecast evaluation is reported via the relative change of the score of the proposed model $\Bar{S}$ to a benchmark $\Bar{S}^{ref}$ via skill scores. If the perfect score is zero, as in the cases considered here, then the skill score is
\begin{equation}
	 \text{skill} = 1-\frac{\Bar{S}}{\Bar{S}^{ref}}
\end{equation}
and in the following the terms skill score, percentage improvement, and relative change are used interchangeably. Bootstrap re-sampling is used as a simple non-parametric method for estimating the significance in forecast improvement~\cite{Messner2020EvaluationView}.
Forecast target times are sampled with replacement with a length equal to the original length of the set, and then skill scores are calculated. This process is repeated a large number of times until the sampling variation of the result is determined, which are then typically presented via boxplots~\cite{Messner2020EvaluationView}.

\subsection{Daily Peak Intensity Evaluation}

\begin{figure}[t]
\centering
\includegraphics[width=\columnwidth]{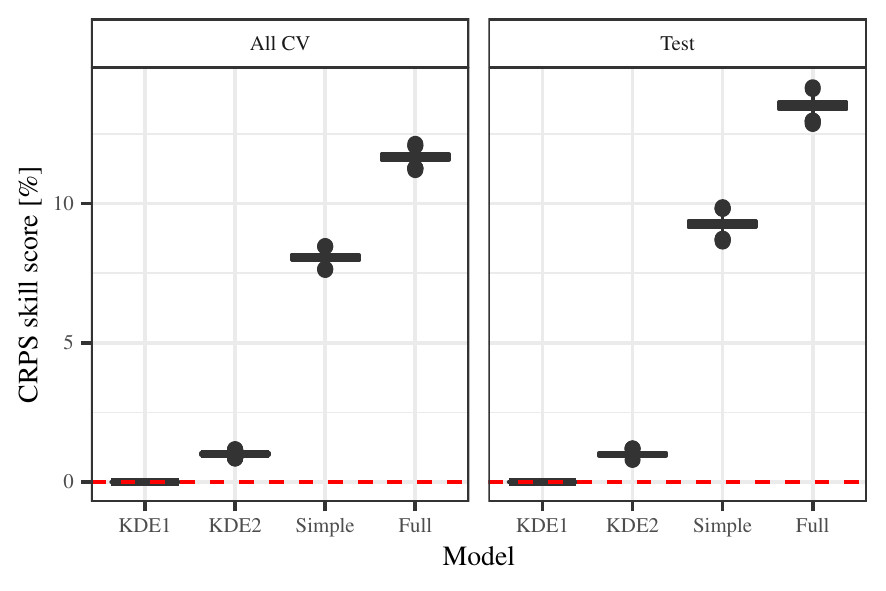}
\caption{Skill scores averages of the peak intensity forecasts at the household level of the network relative to KDE1, where the Full model is a component of the Fusion model. The sample distribution is found via bootstrap averages. The peak intensity forecasts at the household level are as skilful as those at the aggregate levels \deleted{(Figure~[X])} relative to the benchmark.}
\label{fig: boot_peaki_sm}
\end{figure}

\deleted{Although it is perhaps difficult to generate skilful forecasts of the peak timing at the household and lowest voltage levels, it is a different story for the peak intensity.} \replaced{In}{in} Figure~\ref{fig: boot_peaki_sm} the skill of the peak intensity forecasts is demonstrated at the household level against three benchmarks, an unconditional KDE, KDE conditional on day type (weekday/weekend), and a simple GAMLSS model, based on autoregressive features for the location parameter only. The full model is used as a component in the fused forecasts, and as you can see on average results in improved forecasts of over 15\% relative to the most simple benchmark in testing. This validates the motivation for using a bespoke model for predicting the peaks alone. \deleted{Again, as shown in Figure~[X],} There is some variability in the skill between households and this is skewed towards improved skill\added{, as detailed in \cite{AMIDiNe_LV_SupMat2022} and discussed in Section \ref{sec:res_summary}}. Some households \replaced{show}{showing} over a 50\% improvement in CRPS in cross-validation and testing compared to the unconditional KDE estimate and few showing negative skill below -5\%. The improvement between the two GAMLSS models validates the inclusion of the smooth day of year, day type (weekday/weekend), and empty house features.

\deleted{At the aggregated level of the network the skill scores on a per-node basis are shown in Figure~[X]. We first note the the increased sample variation of the skill scores due to smaller size of the daily time-series. However, even given this uncertainty the positive skill of the forecasts is still clear. Interestingly,} The skill scores are similar across all levels of the network from Feeder to Primary Substation, in contrast to peak timing and half-hourly forecasts which are generally more skilful at higher levels of aggregation than lower. \added{The breakdown by individual nodes is presented in \cite{AMIDiNe_LV_SupMat2022} and we find that skill is positive and statistically significant in the majority of cases on the test data, but also note that the sample variation is greater due to the reduced data volume. This is also reflected in how the nodes where Fusion shows the greatest/least skill are not the same between the cross-validation and test results, although the overall behaviour is similar.}




\subsection{Daily Peak Timing Evaluation}

In this subsection we evaluate in more detail the weights used in the fusion forecast, $w_{d,h}$, which is defined as the probability of the daily peak at each node occurring in each discrete block of energy throughout the day. At the aggregate levels, the skill scores of the discrete time-to-event probability forecasts, where the reference is seasonal climatology, are \replaced{range from 0\% (for four feeders) to over 20\% for the Primary Substation. Detailed results may be found in \cite{AMIDiNe_LV_SupMat2022}. There is a clear trend for greater improvement at higher levels in the hierarchy.}{shown in Figure~[X]. The first thing to note is that clearly big improvements in skill are possible at the more aggregated levels for the peak timing prediction.} This is because the peak timing is less variable and more smoothly dependent on seasonal, which is easily modelled. Whereas, even at the feeder level during testing some of the nodes have similar or less skill than the benchmark, although there is a skew generally for positive skill scores. At the feeder levels the time of peak is clearly dependent on more complex behaviours than can be described by time of day and day of year. Additionally, the general change in skill between cross-validation and testing is worth further investigation at all the levels; ideally more data would be available when learning complex seasonal interactions between day of year and time of day which may be leading to overfitting.


This problem is also evident at the smart meter level.
The time-to-event based forecast is only marginally more skilful in testing than the benchmark with a CRPS skill score of under 0.5\%. This highlights the difficulty in predicting the peak timing at the household level and that the predictions of the time-to-event based model are close to seasonal climatology on average. \replaced{The variability in skill has been investigated, shown in \cite{AMIDiNe_LV_SupMat2022}. There is variation in skill between households, which ranges from $\pm$10\%; some households are apparently more predictable than others.}{In Figure~[X] the density of the skill scores across the household level are plotted. There is a node dependent variability in skill, and clearly some households are more predictable and seasonally dependent than others.}



Teasing out relationships between node type, the skill of the timing probability forecasts, and drivers of predictability is an interesting aspect of future work. Stakeholders could use this information to gauge locations for the provision of flexibility in the LV network. Finally, further analysis of the forecasts could give an understanding as to which nodes are likely to reach their daily peak at the same time as the more aggregated nodes higher up the network. This information could be valuable for revealing which households or nodes to leverage for peak demand shifting via (for example) time-of-use tariffs.

\subsection{Forecast Fusion Evaluation}

In the following subsections we first evaluate the forecast fusion methodology. To this end, we evaluate the forecasts using CRPS skill scores averaged over all time periods, and also inspect the calibration directly. \replaced{However}{however}, to demonstrate the improved forecast skill for the daily peak demand we also retrospectively select periods where the daily peak demand is recorded and evaluate the fused forecast averaged over these time periods only. Due to the double penalty effect~\cite{Haben2014AConsumption}, an advanced (and smoother) forecast might produce improved skill on average compared to a benchmark, but fail predict peak demand well. Therefore, the structure of this evaluation is aimed to demonstrated the skill of the forecast on average and during the daily peak demand.

\subsubsection{Aggregate Levels}

\begin{figure}[t]
\centering
\includegraphics[width=\columnwidth]{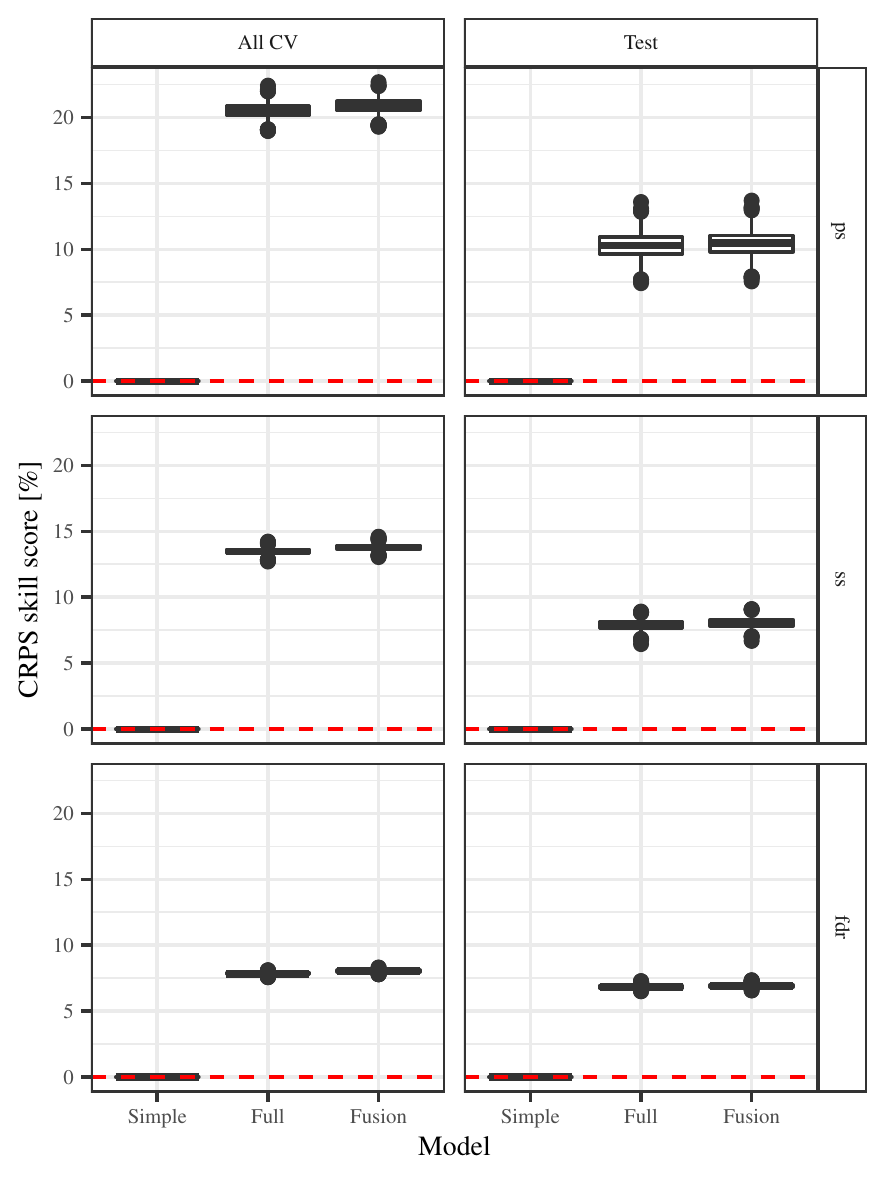}
\caption{Skill scores averages of the three half-hourly forecasting methods employed at the aggregated levels of the network relative to the `Simple' model. The sample distribution is found via bootstrap averages, where all available samples are included at the the primary substation (\texttt{ps}), secondary substation (\texttt{ss}), and feeder (\texttt{fdr}) levels.}
\label{fig: boot_hh_agg}
\end{figure}

At the aggregate levels, the two advanced forecasting methods show similar improved skill over the benchmark at the primary substation (\texttt{ps}), secondary substation (\texttt{ss}), and feeder (\texttt{fdr}) levels in the network. This is true for both cross-validation and testing, as shown in Figure~\ref{fig: boot_hh_agg}. So we can conclude that the forecast fusion method is at least as skilful as the full half-hourly model. This is encouraging since the latter is a key component in the fused forecast. More generally, we can see the skill improvements possible from adding seasonal features and interactions in the regression model, as evidenced by the $\approx$10\% improvement in forecast skill during testing across the aggregated network compared to the simple autoregressive benchmark.

\begin{figure}[t]
\centering
\includegraphics[width=\columnwidth]{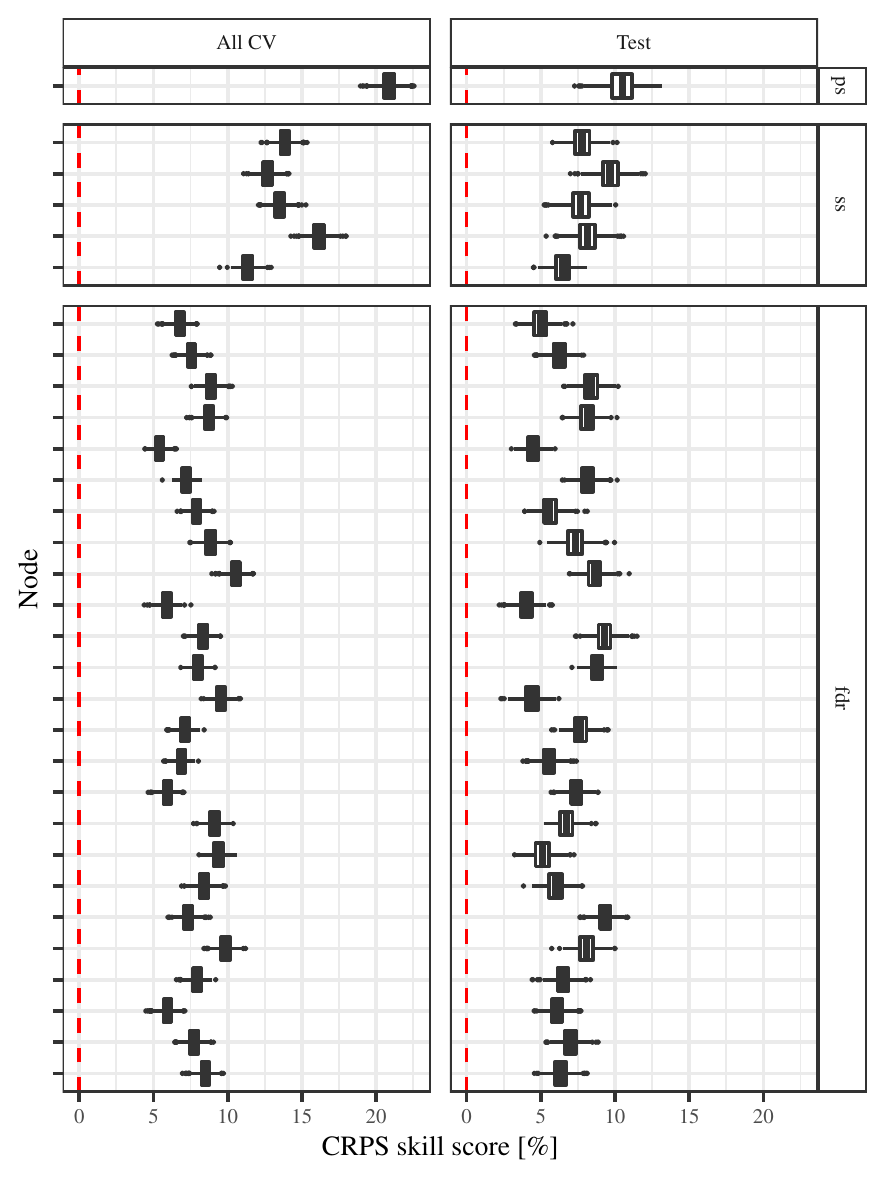}
\caption{Skill scores averages of the Fusion forecasting method at the aggregated levels of the network relative to the `Simple' model. The sample distribution is found via bootstrap averages, where all available samples are included at each node.}
\label{fig: boot_hh_aggid}
\end{figure}

Rather than metrics at each level of the network, in Figure~\ref{fig: boot_hh_aggid} the skill scores of the Fusion method are plotted at each single node of the aggregated network. Clearly there is variability in improvement between nodes, especially at the feeder level which is hidden in Figure~\ref{fig: boot_hh_agg}. This plot also emphasises that during cross-validation the improvement is increasing with the aggregation (or voltage) level. However, during testing the skill scores are far closer between the levels which could indicate that the testing data is more difficult to predict at the aggregate levels, the benchmark models are improved relative to the advanced models with more training data, or the advanced models are over-fitting slightly, or a combination of all three. 

\begin{figure}[t]
\centering
\includegraphics[width=\columnwidth]{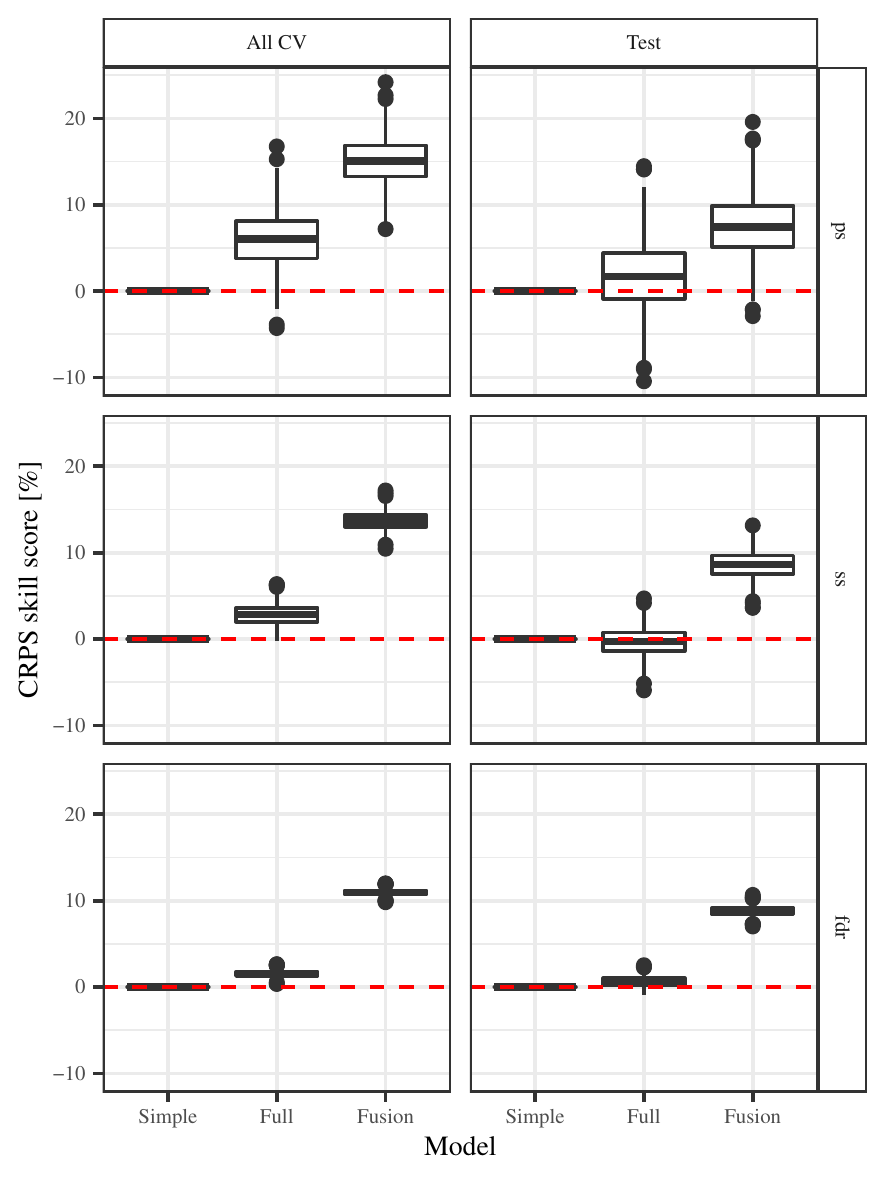}
\caption{Skill scores averages of the three half-hourly forecasting methods employed at the aggregated levels of the network relative to the `Simple' model. The sample distribution is found via bootstrap averages, where only samples which correspond to the daily peak demand are included at the the primary substation (\texttt{ps}), secondary substation (\texttt{ss}), and feeder (\texttt{fdr}) levels}
\label{fig: boot_hhpk_agg}
\end{figure}

If we evaluate the forecasts only during the periods when the daily peak demand is recorded the skill of the three forecasting methods looks very different. Figure~\ref{fig: boot_hhpk_agg} shows that the full model is similar in skill during testing to the benchmark. In fact, during the data exploration and tuning of the models, it proved difficult to find a suitable feature set which performed equally well to the benchmark model during the peak half-hours, as evidenced in the supplementary material~\cite{AMIDiNe_LV_SupMat2022}. However, the Fusion method is again $\approx$10\% better at forecasting the daily peak during testing across the aggregated network compared to the benchmark. This is a key result of the paper and shows that by fusing a bespoke forecast of the daily peaks to a state-of-the-art half-hourly forecasts, it is possible 
to achieve skilful forecasts on average \textit{and} during the daily peak demand. 
\replaced{We have also investigated the variation in forecast performance across individual nodes of the aggregated network. While there is again variability between nodes, all show improvement from forecast fusion of between 5\% and 15\%. Bootstrap skill scores by node are illustrated in \cite{AMIDiNe_LV_SupMat2022}.}{In Figure~[X] the skill scores during daily peak demand of the Fusion method are plotted at each single node of the aggregated network. Clearly there is again some node variability in the improvement. However, during testing the skill scores all converge on $\approx$10\%.}

\begin{figure}[t]
\centering
\includegraphics[width=\columnwidth]{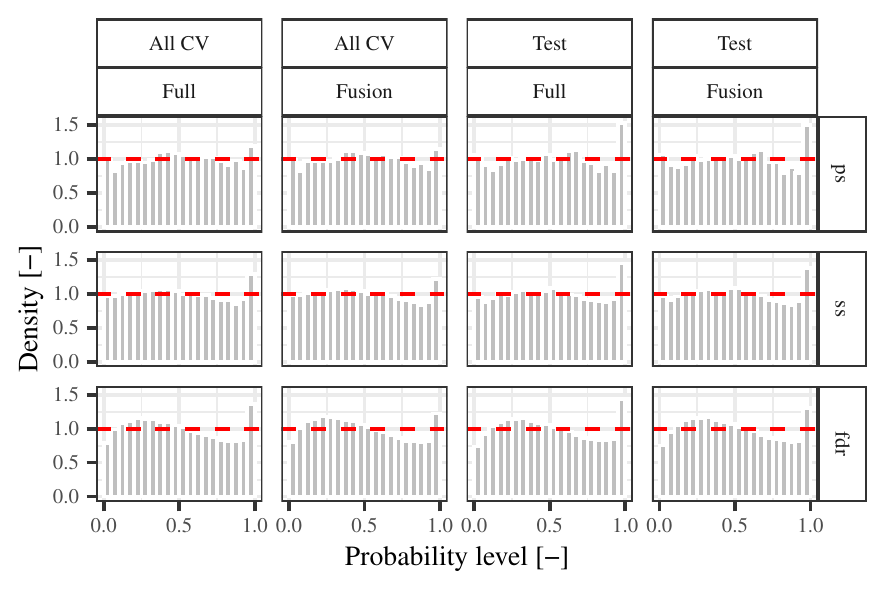}
\caption{PIT histograms of two forecast models at the aggregated levels of the network. Note that except from the primary substation (\texttt{ps}) level, these histograms show the average calibration of all the nodes. The fused forecast is similarly calibrated at all levels and across both data partitions.}
\label{fig: pit_hh_agg}
\end{figure}

The calibration of the advanced GAMLSS and fusion models is shown in Figure~\ref{fig: pit_hh_agg} where essentially the average calibration at the aggregated levels is shown. Importantly, the Fusion method is at least as well calibrated as the Full model, if not marginally better calibrated. Clearly, the right tail of the distribution could be improved at most of the levels to account for large peaks in demand. However, this is reserved for future work. Additionally, the calibration of both models at the feeder level is relatively poor on average. Using distribution free regression approaches might be beneficial here as well as accounting for holiday and special events. The key result here is that the forecasts are reasonably well calibrated and the fusion methodology did not introduce calibration issues via the linear combination, an issue that is widely discussed in forecast combination~\cite{Gneiting2013CombiningDistributions}.

\subsubsection{Household Level}

\begin{figure}[t]
\centering
\includegraphics[width=\columnwidth]{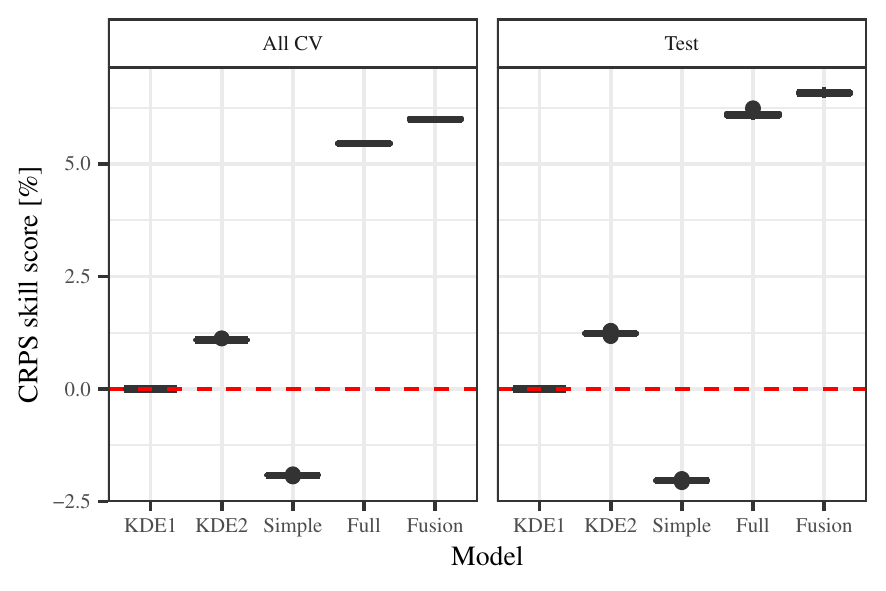}
\caption{Skill score averages of the five half-hourly forecasting methods employed at the household level of the network relative to KDE1. The sample distribution is found via bootstrap averages, where all available samples are included. 
}
\label{fig: boot_hh_sm}
\end{figure}

Recall that there are three benchmarks for household level forecasts, two variations on KDE and a very simple GAMLSS model based only on time-of-day. An interesting result, shown in Figure~\ref{fig: boot_hh_sm} is that using autoregressive and seasonal terms it is possible to achieve better skill than simple benchmarks, by 3--4\% during testing in this case study. As well, the Fusion methodology is marginally better than the advanced GAMLSS model in both cross-validation and testing. The Simple GAMLSS model is the worst performing out of the methods tested. At some nodes (8 out of 742) the Full model failed to converge and the performance of the resulting forecasts was very poor during cross-validation. Further inspection revealed issues at these households such as structural changes in the time-series and so on. At these 8 nodes, the Simple model was therefore used in place of the Full. Additional detail on this can be found in the supplementary material~\cite{AMIDiNe_LV_SupMat2022}.

\replaced{As one would expect, the relative performance of the benchmark and Fusion methods varies between households.}{The variability of the skill scores on a per household basis is shown in Figure~[X], where the density of the household level skill scores is plotted.} Unlike the aggregated level where there was consistent improvement at all nodes for the proposed Fusion methodology, at the household level \replaced{forecast fusion provides improvement for 80\% of households relative to KDE1, and 70\% relative to the sophisticated Full GAMLSS benchmark.}{there are some nodes where the KDE1 benchmark is more skilful.} This is due to the diversity of behaviours at the household level and perhaps at some nodes the model is imply over-parameterised given the information in the time series which suggests perhaps boosting or regularisation would be beneficial in the model fitting if computationally feasible. \replaced{However}{however}, the density is clearly skewed toward improvement as you would expect from Figure~\ref{fig: boot_hh_sm}. \added{Supplementary results on this topic are provided in \cite{AMIDiNe_LV_SupMat2022}.}

\begin{figure}[t]
\centering
\includegraphics[width=\columnwidth]{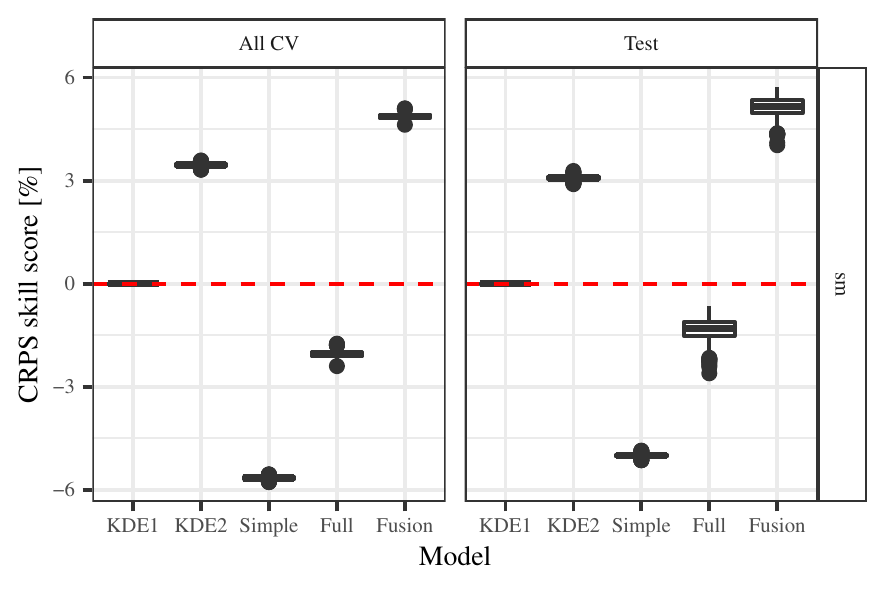}
\caption{Skill scores averages of the Fusion forecasting method at the household level of the network during peaks relative to KDE1. The sample distribution is found via bootstrap averages, using only samples which correspond to the daily peak demand are included at each node.}
\label{fig: boot_hhpk_sm}
\end{figure}

Again, when evaluated during the periods when the daily peak demand is recorded the ranking of the different forecasters is very different. As shown in Figure~\ref{fig: boot_hhpk_sm}, the simple and advanced GAMLSS models are now significantly worse than even the simple time-of-day KDE benchmark by $\approx$1-5\% during testing, which demonstrates the double penalisation effect reported in the literature. However, the fusion model remains the most skilful forecast, and is significantly better than the more advanced KDE model averaged over both cross-validation and testing. Although the skill score is not as large as the aggregated levels detailed above, there is still a large and significant improvement from the full model, which is one of the inputs to the fused forecast, of approximately 6\% during testing.

\begin{figure}[t]
\centering
\includegraphics[width=\columnwidth]{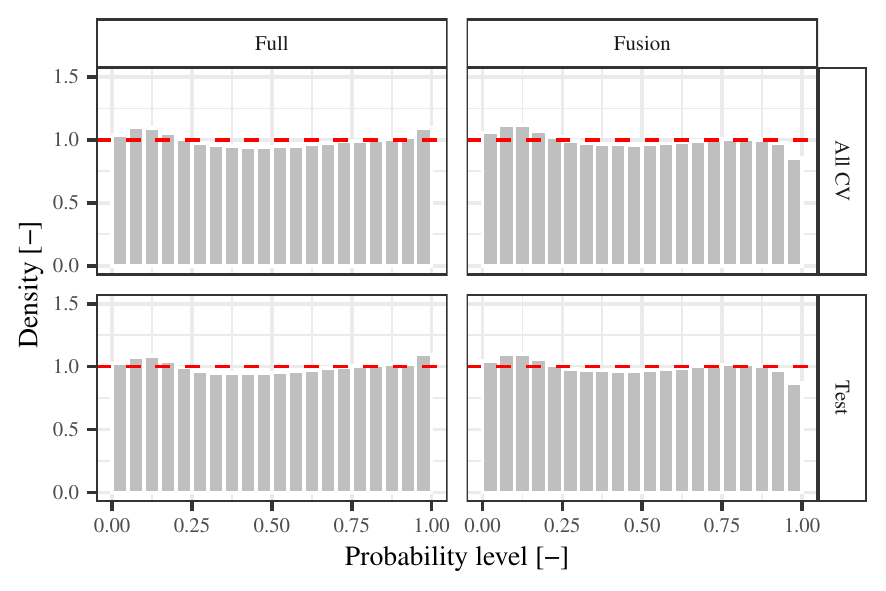}
\caption{PIT histograms of two forecast models at the household level of the network, which show the average calibration of all the nodes. The fused forecast is similarly calibrated at all levels and across both data partitions, except for the right tail of the two methods.}
\label{fig: pit_hh_sm}
\end{figure}

In terms of calibration, the PIT histogram for all the households is shown in Figure~\ref{fig: pit_hh_sm}. Although it is not possible to distinguish individual nodes in this case the plot shows that both the advanced GAMLSS and forecast fusion method are reasonably well calibrated across this level of the hierarchy, with some evidence of over-confidence. however, given the nature of smart meter demand and that we are using a parametric assumption for the predictive distribution, the calibration is better than expected. There are some differences between the calibration of the GAMLSS and Fusion forecast now however, with the right tail of the distribution going from too narrow to too wide on average. This indicates a possible area of improvement for the forecasts.

\subsection{\added{Results Summary and Discussion}}
\label{sec:res_summary}
\added{
    To summarise, at aggregated and household levels, fusion of peak and conventional forecasts provides a significant improvement in CRPS relative to both simple and advanced benchmarks methods during peak hours. The improvement relative to advanced methods ranges from 6\% to 9\%. This comes with no penalty to overall performance, which is also improved but only marginally. Average CRPS is reported for each aggregation level in Table~\ref{tab:crps_summary}, though we refer the interested reader to the earlier figures and supplementary material where we verify that this improvement is largely consistent across individual households, feeders and substations, which differ in size and variability.
}

\begin{table}[t]
\centering
\caption{\added{Summary of the mean CRPS from the test set for the Simple and Advanced (Adv.) half-hourly forecasts, and Fusion forecasts. Scores are presented for All periods and Peak periods. Skill scores are for the Fusion forecasts relative to the Advanced benchmark.}}
\begin{tabular}{l|l|rrr|r}
  \hline
 & Aggregation & Simple & Adv. & Fusion & Skill \\ 
  \hline
\multirow{4}{20pt}{All Time} & Primary & 7.04 & 6.32 & 6.31 & 0.2\% \\ 
   & Secondary & 1.92 & 1.77 & 1.77 & 0.2\% \\ 
   & Feeder & 0.69 & 0.64 & 0.64 & 0.0\% \\ 
   & Household & 0.08 & 0.08 & 0.08 & 0.4\% \\ \hline
  \multirow{4}{20pt}{Peaks Only} & Primary & 8.94 & 8.80 & 8.30 & 5.7\% \\ 
   & Secondary & 3.08 & 3.09 & 2.81 & 9.0\% \\ 
   & Feeder & 1.54 & 1.53 & 1.40 & 8.2\% \\ 
   & Household & 0.47 & 0.47 & 0.44 & 6.0\% \\ 
   \hline
\end{tabular}
\label{tab:crps_summary}
\end{table}

\begin{figure}[t]
\centering
\includegraphics[width=\columnwidth]{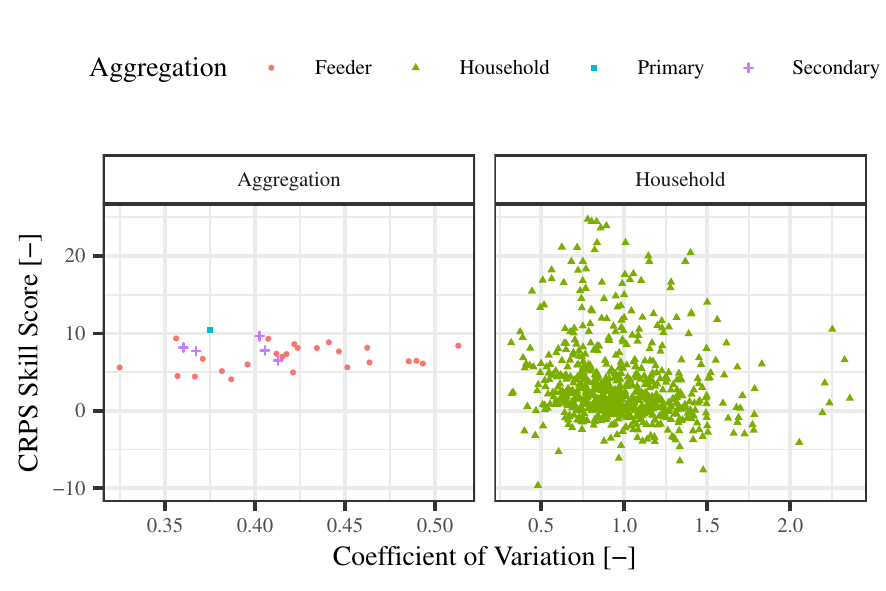}
\caption{Forecast skill of the fusion forecast relative to benchmark (KDE1 for households, Simple half-hourly forecast for aggregations) against coefficient of variation for each substation, feeder and household.}
\label{fig:skill_vs_variation}
\end{figure}

Finally, we have investigated whether there is any relationship between the variability of load (at substations, feeders and households) and forecast improvement of the fusion method relative to benchmarks. We have compared the skill scores with the coefficient of variation \deleted{(standard deviation over mean)} for all individual substations, feeders and households, illustrated in Figure~\ref{fig:skill_vs_variation}. There is no apparent relationship between variability and skill for any aggregate level, which all have positive skill. Of course we have few examples of primary and secondary substations, but skill at these levels is comparable to individual feeders. For households, however, we observe a negative correlation between variability and forecast skill, although there is a large amount of variation, and positive skill for 80\% of households. Furthermore, only households with a relatively low coefficient of variation exhibit very high forecast skill. This highlights the importance of considering forecast skill for individual households, as `average' performance across multiple households will mask this variation in forecast performance. \added{For most applications, we believe that significant improvement at 80\% of households is more than sufficient to justify a slight increase in the complexity of the forecasting process.}

\section{Conclusions}
\label{sec:conculsions}

Forecasting methods that are effective across all voltage levels of distribution networks will be essential as Distribution Network Operators take on new responsibilities for managing energy balancing and ancillary services. This paper presents a novel approach to probabilistic load forecasting that addresses deficiencies of existing methods caused by peak loads, and is shown to improve forecast skill across distribution networks from household to primary substation level by as much as 10\% overall, and more during peaks. The skill of forecasts during peaks is particularly important in many use-cases, including network constraint management, peak shaving, and battery and demand response scheduling.

The approach we propose combines forecasts of daily peak timing and intensity with conventional load forecasts. By forecasting peaks specifically, we \replaced{compensate for}{avoid} the tendency of conventional methods to be too smooth and under-forecast peaks. Probabilistic forecasts are combined or `fused' using a simple weighting scheme inspired by the more general practice of data fusion. A comprehensive case study based on open data is presented where we find that while sophisticated methods for conventional forecasts may provide skill overall with respect to competitive benchmarks, they add little value during peaks. Fusion of conventional forecast with \deleted{with} a peak forecast \added{marginally} improves performance overall, and greatly improves performance during peaks. \added{Average improvement during peaks ranged from 6\% at household and Primary substation level, and 8--9\% at Feeder and Secondary substation level in our case study.}

\added{Additionally, we have proposed a method for producing parametric density forecasts at the household level based on the Generalised Beta-Prime distribution and the GAMLSS framework. This is in contrast to non-parametric methods that have dominated the literature to date and are far less parsimonious. The proposed method produced an average improvement of approximately 5\% relative to methods based on Kernel Density Estimation across the 742 households in our case study.}

However, \replaced{forecasting capabilities require further development}{forecasting capability requires further develop} to meet the expected future needs of \glspl{dso}. Not least, consideration should be given to embedded generation, storage, and demand response. Furthermore, to be of maximum practical use, forecasting models should be applicable to feeders/substations they have not been trained on\added{, known as `global' forecasting models as proposed in \cite{Grabner2022ANetworks}}. As distribution networks feature tens-of-thousands of LV feeders, the use of domain adaptation via transfer learning would take a pre-trained model and adapt it to any feeder given minimal adjustment, but certainly not retraining. Development of forecasting models that are adaptive to track changes in load behaviours, structural breaks in particular, should also be considered.
Another aspect to consider and potentially exploit is the hierarchical nature of electricity demand. Encoding this structure in forecasting models can help improve accuracy and enable more coordinated decisions at different levels of the network.







\section*{Acknowledgements}

The authors would like to thank Stephen Haben for many insightful discussions on this topic which have informed this work and UK Power Networks for provision of the Low Carbon London dataset, available at https://data.london.gov.uk/dataset/smartmeter-energy-use-data-in-london-households, or the pre-processed version used here \cite{AMIDiNe_LV_SupMat2022}. This study is fully reproducible, all data and R code associated with this work are available in \cite{AMIDiNe_LV_SupMat2022}. The authors were funded by the EPSRC project Analytical Middleware for Informed Distribution Networks (AMIDiNe, EP/S030131/1) and the Innovation Fellowship held by JB (EP/R023484/1 and EP/R023484/2).




\bibliographystyle{elsarticle-num} 
\bibliography{references_JB}





\end{document}


\maketitle

This document contains supplementary results to support the analysis presented in the main paper. These are primarily breakdowns of forecast evaluation by individual secondary substations, feeders, and households. Their purpose is to illustrate how improvements in forecast performance is distributed across individual households/feeders/substations. For substations and feeders, forecast improvement is very consistent and of a similar magnitude across most substations/feeders. The Fusion method improves performance at 80\% of households compared to KDE1, and 70\% of households compared the `Full' half-hourly benchmark GAMLSS model.

\begin{figure}[t]
\centering
\includegraphics[width=\columnwidth]{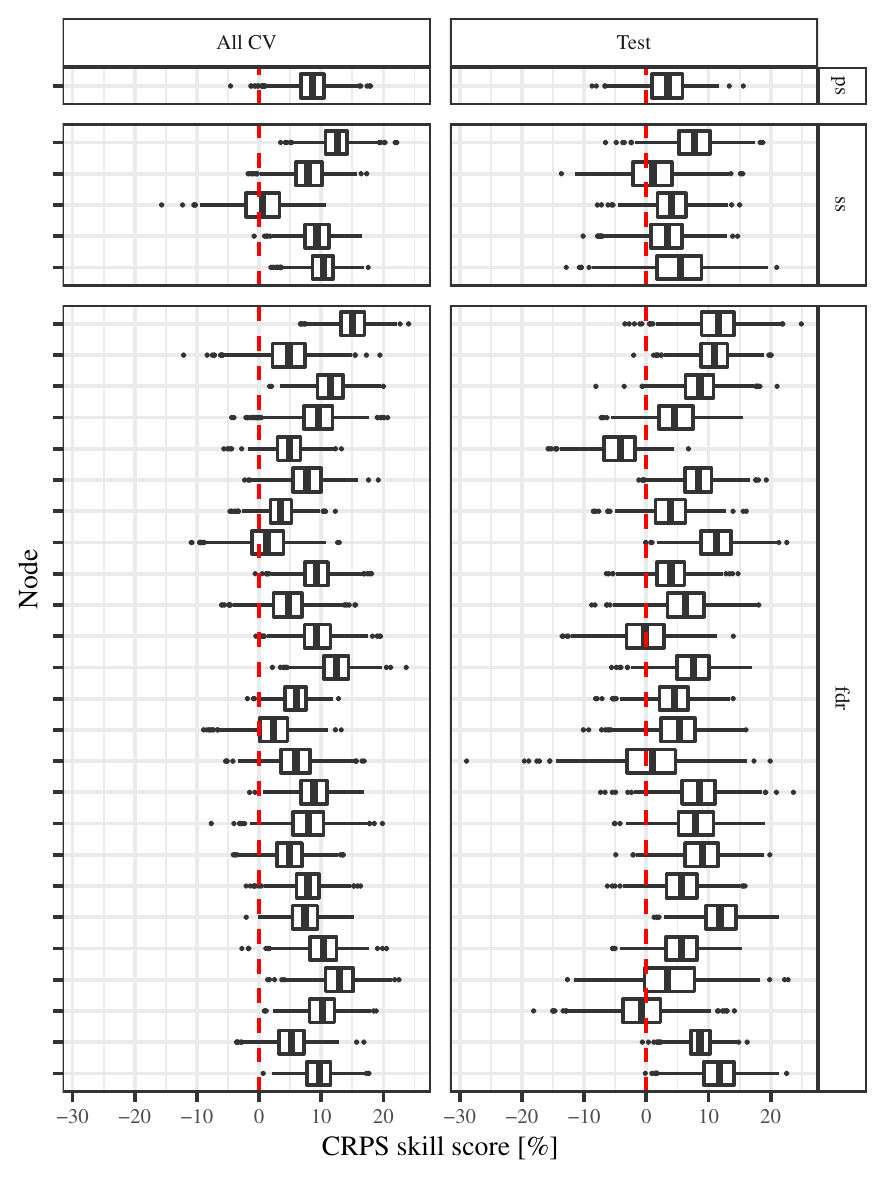}
\caption{Skill scores averages of the advanced peak intensity forecasts at the aggregated nodes of the network, where unconditional climatology is the reference model. The sample distribution is found via bootstrap averages, and is naturally wider compared to the other evaluation plots because of the reduced sample size at each node. The forecast skill is posttive and similar around the aggregated network on average.}
\label{fig: boot_peaki_agg}
\end{figure}

\begin{figure}[t]
\centering
\includegraphics[width=\columnwidth]{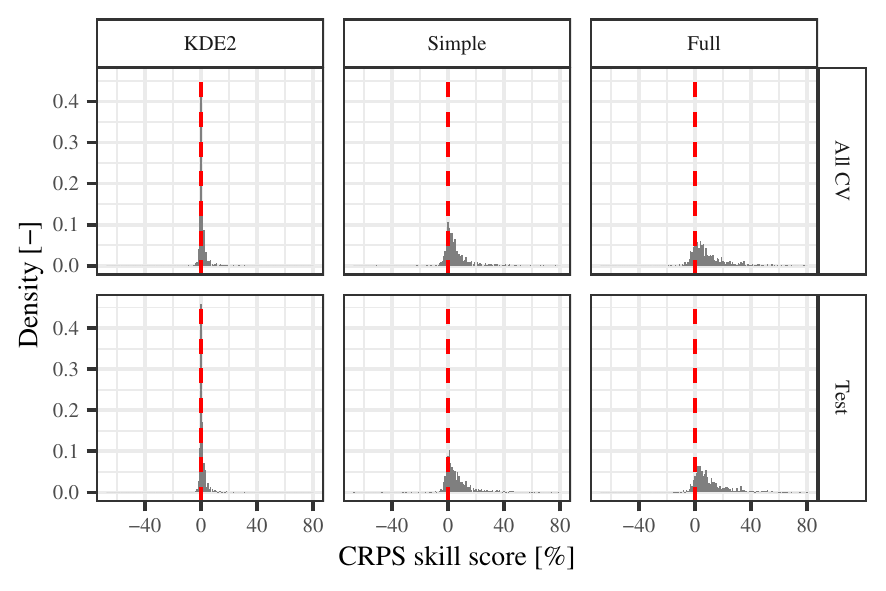}
\caption{Density of the skill scores at the household level where the reference model is KDE1. There is variable improvement across households in terms of predicting the daily intensity, a clear positive skew for both \texttt{GAMLSS} models.}
\label{fig: dens_peaki_sm}
\end{figure}

\begin{figure}[t]
\centering
\includegraphics[width=\columnwidth]{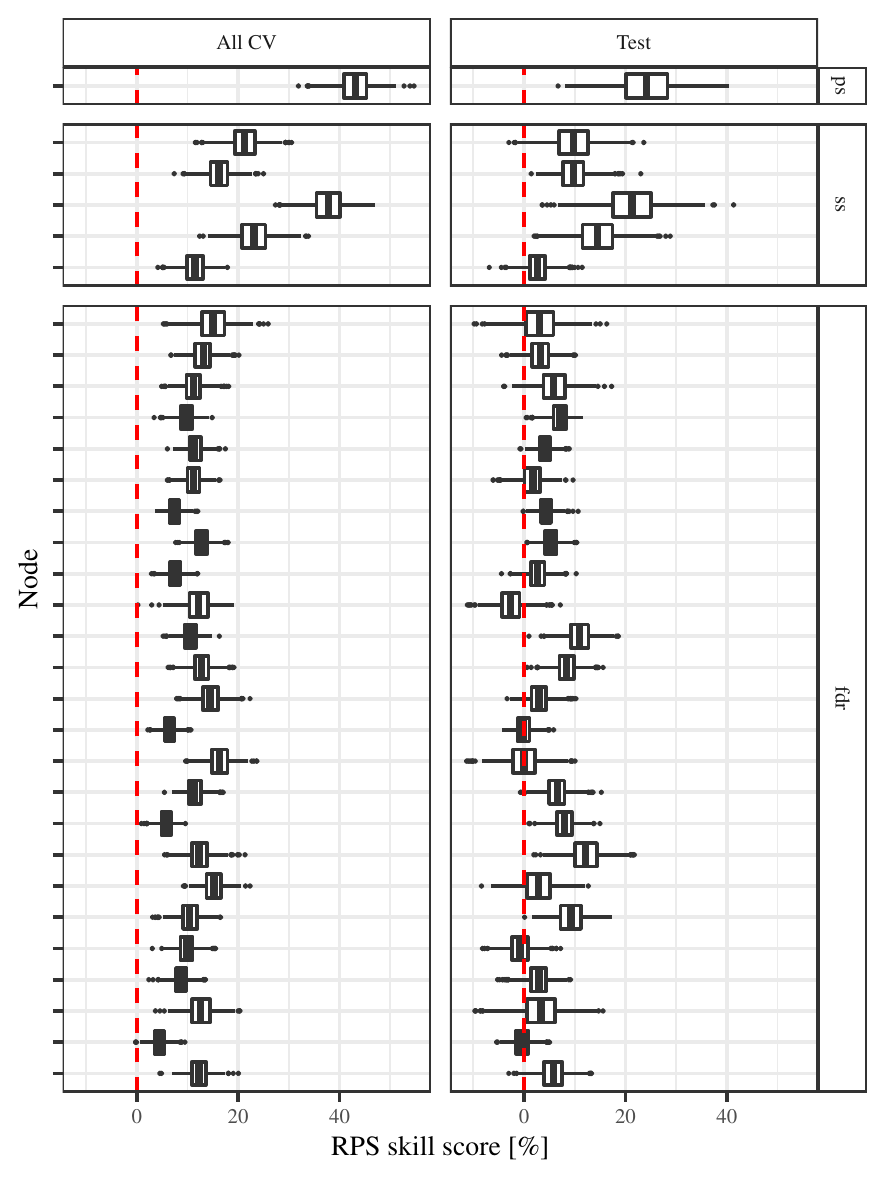}
\caption{Skill scores averages of the time to peak forecasts at the aggregated nodes of the network, where seasonal climatology is the reference model. The sample distribution is found via bootstrap averages. As expected, it becomes more difficult to predict the peak timing at the lower voltage levels.}
\label{fig: boot_peakt_agg}
\end{figure}

\begin{figure}[t]
\centering
\includegraphics[width=\columnwidth]{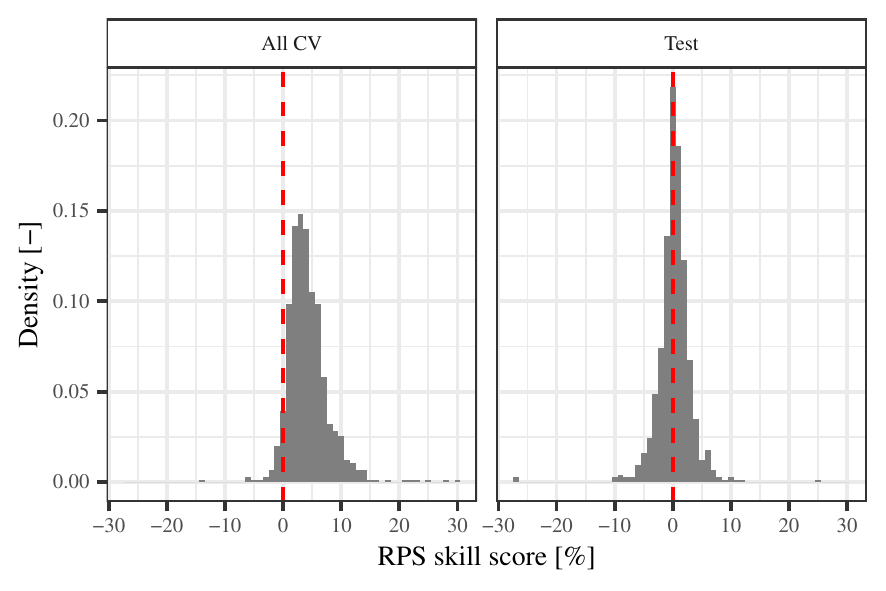}
\caption{Density of the skill scores for peak timing forecasts at all the smart meter nodes where the reference model is seasonal climatology. There is variable improvement around the households in terms of predicting the daily peak timing. In testing the average skill is close to 0\%.}
\label{fig: dens_peakt_sm}
\end{figure}

\begin{figure}[t]
\centering
\includegraphics[width=\columnwidth]{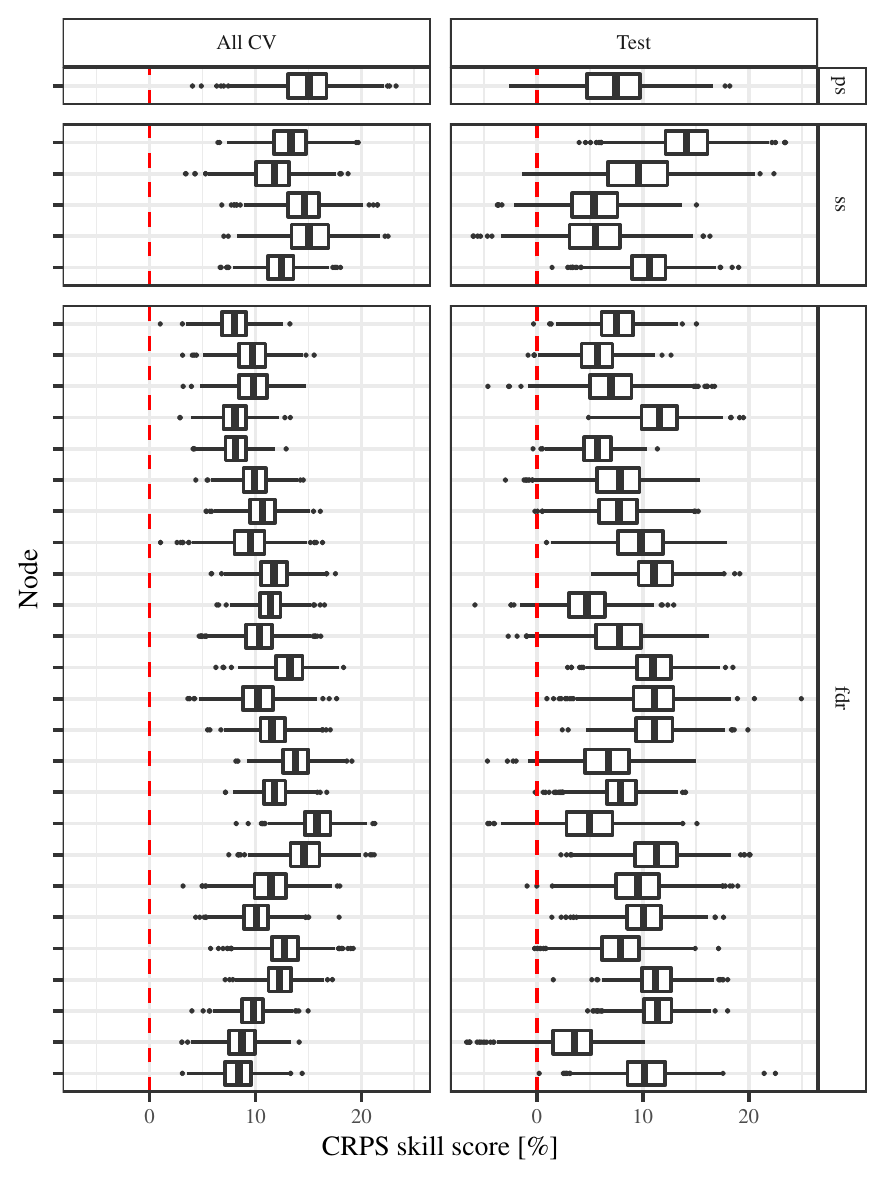}
\caption{Skill scores averages of the Fusion forecasting method at the aggregated levels of the network relative to the Simple model. The sample distribution is found via bootstrap averages, where only samples which correspond to the daily peak demand are included at each node.}
\label{fig: boot_hhpk_aggid}
\end{figure}

\begin{figure}[t]
\centering
\includegraphics[width=\columnwidth]{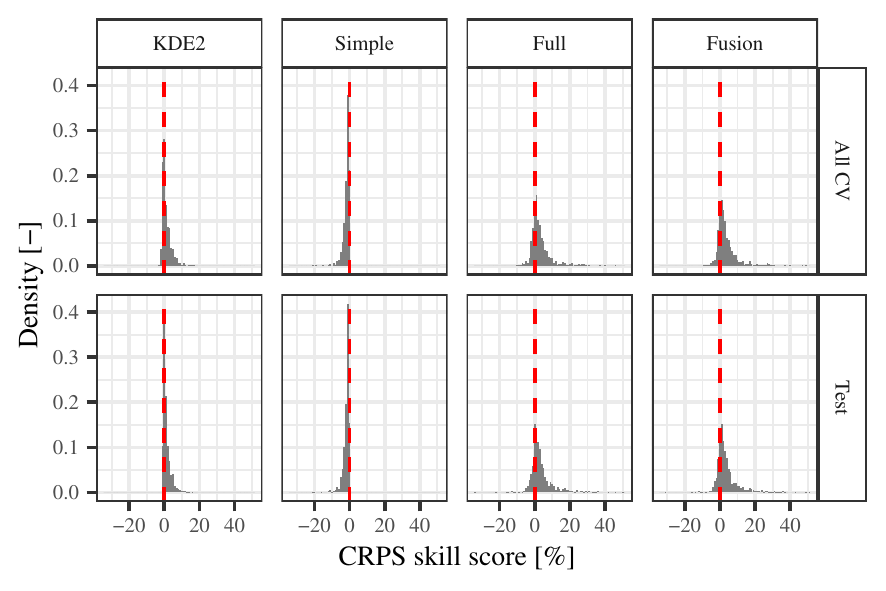}
\caption{Density of the skill scores at all the smart meter nodes relative to KDE1. Although there is variable improvement around the households the distribution is clearly skewed in favour of the \texttt{fusion} forecasts in both cross validation and testing.}
\label{fig: dens_hh_sm}
\end{figure}